\documentclass[aps,onecolumn,amsmath,amsfonts,nofootinbib,superscriptaddress,
  prd,eqsecnum,secnumarabic]{revtex4}
\usepackage{color,graphicx,dcolumn,bm,multirow,float,hyperref,ulem}
\usepackage[utf8]{inputenc}
\usepackage{array}
\hypersetup{colorlinks=true,linkcolor=blue,citecolor=magenta,filecolor=magenta,
  urlcolor=cyan}

\definecolor{darkmagenta}{rgb}{0.55, 0.0, 0.55}
\definecolor{blue}{rgb}{0.0, 0.5, 0.69}

\begin{document}

\title{Future lepton collider prospects for a ubiquitous composite
  pseudo-scalar}

\author{Alan S. Cornell}
\email{acornell@uj.ac.za}
\affiliation{Department of Physics, University of Johannesburg, PO Box 524,
  Auckland Park 2006, South Africa}

\author{Aldo Deandrea}
\email{deandrea@ipnl.in2p3.fr}
\affiliation{Universit\'e de Lyon, F-69622 Lyon, France: Universit\'e Lyon 1,
  Villeurbanne CNRS/IN2P3, UMR5822, Institut de Physique des 2 Infinis de Lyon}

\author{Benjamin Fuks}
\email{fuks@lpthe.jussieu.fr}
\affiliation{Laboratoire de Physique Th\'eorique et Hautes Energies (LPTHE),
  UMR 7589, Sorbonne Universit\'e et CNRS, 4 place Jussieu,
  75252 Paris Cedex 05, France}
\affiliation{Institut Universitaire de France, 103 boulevard Saint-Michel,
  75005 Paris, France}

\author{Lara Mason}
\email{mason@ipnl.in2p3.fr}
\affiliation{Department of Physics, University of Johannesburg, PO Box 524,
  Auckland Park 2006, South Africa}
\affiliation{Universit\'e de Lyon, F-69622 Lyon, France: Universit\'e Lyon 1,
  Villeurbanne CNRS/IN2P3, UMR5822, Institut de Physique des 2 Infinis de Lyon}

\date{\today}

\begin{abstract}
Composite Higgs models feature new strong dynamics leading to the description
of the Higgs boson as a bound state arising from the breaking of a global
(flavour) symmetry. These models generally include light states generated by the
same dynamics, the detection of which may present the first observable signs of
compositeness. One such state is a pseudo-scalar boson resulting from the
breaking of a $U(1)$ symmetry common to most composite setups, and whose hints
are expected to be visible through low-mass resonance searches at present and
future hadron and lepton colliders. In this work we study the phenomenology of
this pseudo-scalar field. We show that, for a light state, bottom quark loop
effects dominantly impact the production cross section and considerably modify
the decay pattern. Moreover, we make a case for targeted low-mass analyses at
future lepton colliders, with an emphasis on high-luminosity machines aiming to
operate at low centre-of-mass energies. We present a simplified outline of a
search for this light pseudo-scalar at one such machine, considering
electron-positron collisions at the $Z$-pole. We focus on a signature arising
from the pseudo-scalar decay into a pair of hadronic taus and a production mode
association with a pair of leptons of opposite electric charges, and compare cut and count methods with machine learning methods.
\end{abstract}

\maketitle

%\tableofcontents

%%%%%%%%%%%%%%%%%%%%%%%%%%%%%%%%%%%%%%%%%%%%
%  Section 1: Introduction

\section{Introduction}
One of the foremost goals of the current experimental high energy physics
programme is the search for new resonances. The LHC is in a long shut-down
following its 13~TeV run, preparing to operate at its 14~TeV design energy as well as for a high luminosity (HL-LHC) run. At these higher energies and luminosities, efforts will be focused predominantly on the search for resonances typically heavier than the Higgs boson. However, new physics may still be concealed at lower energies. Proposals for electron-positron colliders designed to be complementary to the LHC have been put forward, including the International Linear Collider (ILC)~\cite{Behnke:2013xla}, with initial energies in the range of 250--500~GeV and ranging up to 1~TeV, the Compact Linear Collider (CLIC)~\cite{Charles:2018vfv}, which could reach up to 3~TeV, as well as the Future Circular Collider (FCC-ee)~\cite{Abada:2019zxq} and Circular Electron Positron Collider (CEPC)~\cite{CEPCStudyGroup:2018ghi}, which will operate around the
$Z$ pole, the $WW$ and $t\bar t$ thresholds, and in a Higgs factory mode.
These machines, with lower centre-of-mass (c.m.) energies than the LHC, are
designed to be `factories' for resonances such as the $Z$, $W$, and
Higgs bosons and the top quark, offering the possibility of precision
measurements of their couplings and related Lagrangian parameters.
They can also be used for targeted low-mass resonance searches, providing a
window to possible Beyond the Standard Model (BSM) physics at lower energy
scales through large integrated luminosities. In this article we investigate the
potential of finding hints for a new light pseudo-scalar $a$ emerging from a
composite Higgs model, which may be accessed at both hadron and lepton
colliders.

Composite Higgs models~\cite{Kaplan:1983fs,Kaplan:1983sm, Dugan:1984hq} describe the Standard Model (SM) Higgs sector in terms of high-scale fundamental gauge dynamics by postulating the existence of a new
strong sector. These models implement gauge and fermionic degrees of
freedom, confining at low energies~\cite{Cacciapaglia:2014uja}. In the following
we describe a class of composite Higgs models featuring fermionic
matter~\cite{Ferretti:2013kya,Barnard:2013zea}, charged under a global symmetry
$G$ and governed by a hyper-colour gauge group $G_{HC}$. The breaking of $G$
then leads to the appearance of resonances that are bound states of the
underlying fermions~\cite{Arbey:2015exa}. Composite Higgs models~\cite{Agashe:2004rs,Contino:2010rs,Bellazzini:2014yua, Panico:2015jxa} are an
attractive class of BSM theories as they provide a solution to the hierarchy
problem inherent to the SM whose Higgs sector is unstable with respect to
quantum corrections. If the Higgs boson is not an elementary scalar but rather a
bound state of strong dynamics, these quantum corrections may only contribute up
to a finite scale, hence stabilising the Higgs field
dynamics~\cite{Contino:2010rs}. Introducing this scale of compositeness is one
of a few options for a natural generation of the Higgs boson mass, and offers an explanation for the scale of electroweak symmetry breaking~\cite{Panico:2015jxa}.

In addition to the breaking of the electroweak symmetry~\cite{Weinberg:1975gm}, occurring at a scale $v\sim 246$~GeV, the global symmetries of the fundamental fermion sector are broken at some condensation scale on the order of 1~TeV~\cite{Cacciapaglia:2019bqz}. The possibilities for global and gauge symmetries within a composite Higgs model, though subject to some constraints, are fairly broad. As such, we allow in our analysis the group structures and symmetries within the theory to vary, considering a spectrum of twelve possible models that have recently been proposed as the most minimal options for a composite high-scale dynamics featuring solely fundamental fermions~\cite{Ferretti:2013kya}. These models, which are strongly coupled in the IR, employ a minimal set of fields and depend on fully computable parameters.
Fermion mass generation is achieved via partial compositeness~\cite{Kaplan:1991dc}; however, given
constraints on asymptotic freedom, this mechanism is limited to the generation
of the mass of the top quark~\cite{Ferretti:2014qta}.  While the global symmetries of the fermions and the hyper-colour groups vary across the models, similarities across composite Higgs models of this nature remain, particularly at the effective low energy scale. In particular, in addition to the existence of a QCD symmetry, there always exists a non-anomalous $U(1)$ symmetry, acting on all species of underlying fermions in the theory~\cite{Cacciapaglia:2017iws, BuarqueFranzosi:2018eaj,Cacciapaglia:2019bqz}.
The first evidence of new physics may therefore arise from direct searches for
the additional light state $a$ produced in association with the Higgs boson and
associated with the breaking of this extra $U(1)$. Such a pseudo-scalar state
moreover features interactions with the SM gauge bosons via the
Wess-Zumino-Witten anomaly~\cite{Witten:1983tw,Wess:1971yu}, which is of
particular phenomenological interest concerning its production at colliders.

In order to have evaded detection up to now, a light pseudo-scalar would need to
be weakly coupled to the SM particles, carrying no colour or electric charge.
Our candidate, the ubiquitous $a$, is considered here to have
a mass between 10 and 100~GeV. This range warrants a particular investigation
given the deficiency of LHC searches in the lower end of this mass range thus
far~\cite{Cacciapaglia:2019bqz}. In the considered set of models, the parameters
of the theory and couplings to other states are fully specified and calculable,
allowing for the construction of a general analysis targeting a new light
scalar. To this aim
we have formulated a new unique implementation of the pseudo-scalar $a$ within
state-of-the-art modeling tools, in order to describe simultaneously a range of
composite Higgs models with a variety of group structures. We have
generalised previous implementations of this class of theories~\cite{%
Cacciapaglia:2017iws,Cacciapaglia:2019bqz}, where only loops of SM top quarks
were contributing to the interactions of the pseudo-scalar with the SM gauge
bosons, and the impact of all lighter quarks was taken to be negligible. On the
contrary, our work includes both top and bottom quark contributions, which are
shown to provide non-negligible effects, particularly for the phenomenology of
very light $a$ bosons. This is significant given the interest in low-mass
resonance searches at the LHC, including di-jet~\cite{Sirunyan:2018ikr,%
Aaboud:2018fzt}, di-muon~\cite{Sirunyan:2019wqq,Aad:2019fac}, di-photon~\cite{%
Mariotti:2017vtv,Sirunyan:2018aui,Aaboud:2016tru} and di-tau~\cite{Aaboud:2017sjh} searches in recent years.
These searches yield poor constraints in the low pseudo-scalar mass regions,
but the associated results rely on older cross section predictions, ignoring
bottom-quark loop effects. The regions of interest are therefore better covered
than initially thought.
In particular, di-tau searches consist of one of the golden channels for the
considered pseudo-scalar, as the corresponding branching ratio is usually quite
large~\cite{Cacciapaglia:2017iws}.
The latter are generally strictly restricted to the high-mass
regime, due to the presence of the large QCD background that proves to be an
obstacle to low mass searches. For this reason we have instead
investigated the potential of future lepton colliders where, following an
overview of possibilities across a range of proposed colliders, an analysis at
a low c.m.~energy of the signal induced by the production of the pseudo-scalar
is presented.

In this manuscript we begin, in Sec.~\ref{sec:chmodel}, with a theoretical motivation, describing composite Higgs models which are built on a theory of fundamental fermions and outlining the specifics of the models to be studied. In Sec.~\ref{sec:U1_theory} we describe the dynamics of the boson $a$, which is the subject of this work, and in Sec.~\ref{sec:anomcouplings}, its anomalous couplings to the SM gauge bosons. Sec.~\ref{sec:lowmasssearches} outlines a possible low mass search strategy at lepton colliders, investigating the production of $a$ at a variety of future experiments, before constructing an analysis for
a future electron-positron collider aiming at operating at the $Z$-pole. We in
particular outline and compare the expectation of a cut and count analysis, described in Sec.~\ref{sec:cutcount}, and a machine learning approach based on gradient boosted trees, described in Sec.~\ref{sec:ml}.
We summarise our work and conclude in Sec.~\ref{sec:conclusion}.

%%%%%%%%%%%%%%%%%%%%%%%%%%%%%%%%%%%%%%%%%%%%
%  Section 2: Theoretical motivation

\section{Theoretical motivation}\label{sec:chmodel}
This work considers a class of models~\cite{Ferretti:2013kya} with a variety of
hyper-colour groups and several of the most minimal cosets $G/H$
characterising the dynamics below the confinement scale. The symmetry breaking patterns in each model are determined by the properties of the underlying fermions. For a given model with
$N_f$ Dirac fermions of the same species, we may only have one of two possible global flavour symmetries $G$, namely $SU(2N_f)$ for a (pseudo-)real fermion representation, or $SU(N_f)\times SU(N_f)$ for a complex fermion representation~\cite{Cacciapaglia:2019bqz}. The chiral symmetry breaking may then follow one of three patterns;
$SU(2N_f)\rightarrow SO(2N_f)$ for a real representation, $SU(2N_f)\rightarrow
Sp(2N_f)$ for a pseudo-real one, or $SU(N_f)\times SU(N_f)\rightarrow SU(N_f)$ in the case of a complex representation~\cite{Ferretti:2016upr}.

In a general composite Higgs model, the mass of the SM fermions is generated
through either four-fermion interactions~\cite{vonGersdorff:2015fta} or partial
compositeness~\cite{Kaplan:1991dc}. The latter presents a need for fermions in
(at least) two different irreducible representations of the unbroken
hyper-colour group $G_{HC}$, leading to a rich spectrum within the
theory~\cite{Witzel:2019jbe}. All models considered in this work thus contain
two species of underlying fermions which we denote $\psi$ and $\chi$ following the
notation of Ref.~\cite{Ferretti:2013kya}, and which belong to different
irreducible representations of $G_{HC}$ and $G$. The first fermion $\psi$ gives
rise to the Higgs boson through the breaking of the associated global symmetry
$G$ into the electroweak (EW) coset $H$, and carries electroweak charges. The
misalignment of the Higgs field then drives the usual electroweak symmetry
breaking process, the mass of the Higgs boson being generated through some
explicit breaking in the global sector~\cite{Cacciapaglia:2015iua}.
The second species of fermion $\chi$ is responsible for partial compositeness,
that proceeds through a mixing of the top quark with a composite operator of the
same quantum numbers~\cite{Contino:2010rs}. The fermion $\chi$ hence carries QCD
colour and hypercharge quantum numbers, and the breaking of the global symmetry
then generates the QCD coset.
The traditionally searched-for spin$-\frac{1}{2}$ vector-like top-partners are
therefore composed of fermions in two representations of the hyper-colour group, of the form $\psi\chi\chi$ or $\psi\psi\chi$, and are labelled chimera baryons.

The presence of fermions in two irreducible representations in the theory means
that there will always exist two $U(1)$ axial symmetries resulting from the full
symmetry breaking pattern, one combination of these being non-anomalous
with respect to the confining hyper-colour group~\cite{Belyaev:2016ftv}. As a
result, one of the numerous pseudo-Nambu-Goldstone bosons always turns out to be
light, {\it i.e.}~lighter than the confinement scale. This contrasts with the
anomalous axial current in QCD.
\begin{table}
 \renewcommand\arraystretch{1.55}
 \setlength\tabcolsep{8pt}
  \begin{tabular}{c|cc|cc|c}
   & $G_{HC}$  & EW and QCD coset & $\psi$ & $\chi$ & $q_\chi/q_\psi$    \\
  \hline\hline
  M1 & $SO(7)$ &
    \multirow{2}{*}{$\frac{SU(5)}{SO(5)}\times\frac{SU(6)}{SO(6)}$} &
    \multirow{2}{*}{$5\times$\textbf{F}} &
    \multirow{2}{*}{$6\times$\textbf{Sp}} & $-5/6$\\
  M2 & $SO(9)$ & & & & $-5/12$ \\
  \hline M3 & $SO(7)$ &
    \multirow{2}{*}{$\frac{SU(5)}{SO(5)}\times\frac{SU(6)}{SO(6)}$}  &
    \multirow{2}{*}{$5\times$\textbf{Sp}} &
    \multirow{2}{*}{$6\times$\textbf{F}} & $-5/6$  \\
  M4 & $SO(9)$ & & & & $-5/3$ \\
  \hline M5  & $Sp(4)$ &
    $\frac{SU(5)}{SO(5)}\times\frac{SU(6)}{SO(6)}$  &
    $5\times$\textbf{A}$_2$ &
    $6\times$\textbf{F} & $-5/3$ \\
  \hline M6 & $SU(4)$ &
    \multirow{2}{*}{$\frac{SU(5)}{SO(5)}\times\frac{SU(3)^2}{SU(3)}$}  &
    $5\times$\textbf{A}$_2$ & $3\times$(\textbf{F},$\overline{\textbf{F}}$) &
    $-5/3$ \\
  M7 & $SO(10)$ & &
    $5\times$\textbf{F}     & $3\times$(\textbf{Sp},$\overline{\textbf{Sp}}$) &
    $-5/12$ \\
  \hline M8 & $Sp(4)$ &
    \multirow{2}{*}{$\frac{SU(4)}{Sp(4)}\times\frac{SU(6)}{SO(6)}$}  &
    $4\times$\textbf{F}$_2$ & $6\times$\textbf{A}$_2$ & $-1/3$  \\
  M9 & $SO(11)$ & &
    $4\times$\textbf{Sp}    & $6\times$\textbf{F},  & $-8/3$ \\
  \hline M10 & $SO(10)$ &
    \multirow{2}{*}{$\frac{SU(4)^2}{SU(4)}\times\frac{SU(6)}{SO(6)}$}  &
    $4\times$(\textbf{Sp},$\overline{\textbf{Sp}}$) &
    $6\times$\textbf{F} & $-8/3$   \\
  M11 & $SU(4)$ & &
    4$\times$(\textbf{F},$\overline{\textbf{F}}$)   &
    $6\times$\textbf{A}$_2$ & $-2/3$ \\
  \hline M12  & $SU(5)$ &
    $\frac{SU(4)^2}{SU(4)}\times\frac{SU(3)^2}{SU(3)}$ &
    4$\times$(\textbf{F},$\overline{\textbf{F}}$) &
    3$\times$(\textbf{A}$_2,\overline{\textbf{A}}_2$) &
    $-4/9$
  \end{tabular}
  \caption{Key features of the models studied in this work and described further
    in Refs.~\cite{Belyaev:2016ftv,Cacciapaglia:2017iws,Cacciapaglia:2019bqz}.
    The first column contains the model naming convention, and the second
    indicates the confining hyper-colour gauge group, followed by the EW and QCD
    cosets (third column). We then provide the irreducible representations of
    the fermions $\psi$ (fourth column) and $\chi$ (fifth column) under the
    coset choice. The final column includes the ratio of the fermion charges
    under the non-anomalous $U(1)$ symmetry.}
\label{tab:models}
\end{table}
In a composite Higgs model we thus expect a low energy spectrum in which the Higgs boson is accompanied by exotic composite scalars and fermions, some of which are ubiquitous to all composite Higgs models and are of the $\langle \psi\psi\rangle, \langle\chi\chi\rangle, \langle\psi\chi\chi\rangle$ or $\langle\psi\psi\chi\rangle$ forms. Notably, the condensation of the underlying fermions also breaks the pervading non-anomalous $U(1)$ symmetry, leading to two massive singlet physical eigenstates that we denote $a$ and $\eta^\prime$, where $a$ is the lightest state. There exists some non-trivial mixing between the two corresponding gauge eigenstates, which depends on the characteristics of the underlying fermionic sector.
The mixing angle $\alpha_{\rm dec}$ reads, in the decoupling limit~\cite{Cacciapaglia:2019bqz},
\begin{equation}
  {\rm cosec}\ \alpha_{\rm dec} = -\
     \sqrt{ 1+ \frac{q_\psi^2N_{\psi}f_\psi^2}{ q_\chi^2N_{\chi}f_\chi^2} } \ ,
\end{equation}
where $q_\psi$ ($q_\chi$) is the charge of the fermion $\psi$ ($\chi$) under the
non-anomalous $U(1)$ symmetry, $N_\psi$ ($N_\chi$) its multiplicity, and
$f_\psi$ ($f_\chi$) its decay constant.
In the decoupling limit that we consider in this work, all other states decouple
so that one solely focuses on the phenomenology of the light pseudo-scalar $a$.
The range of models according to the most minimal choices for the gauge
structure, numbered M1--M12~\cite{Ferretti:2013kya}, are presented
in Tab.~\ref{tab:models} in which we summarise their properties. Each model is
there defined by a hyper-colour symmetry group $G_{HC}$, given together with
the corresponding irreducible representations of the two species of fermions.
The table also specifies the EW and QCD cosets for each model. A great advantage to these models is the computability of all low-energy parameters, which are completely determined by the underlying fermion construction.  The models are
further described in detail in Refs.~\cite{Belyaev:2016ftv,Cacciapaglia:2017iws,Cacciapaglia:2019bqz}.

%%%%%%%%%%%%%%%%%%%%%%%%%%%%%%%%%%%%%%%%%%%%

\subsection{A light $U(1)$ pseudo-scalar}\label{sec:U1_theory}
In order to study the phenomenology of the light pseudo-scalar $a$ at colliders,
we have constructed a new {\sc FeynRules}~\cite{Alloul:2013bka} implementation
of a simplified model describing the dynamics of the $a$ state in a general way.
This allows for the generation of UFO model files~\cite{Degrande:2011ua} that
can be used further within the {\sc MadGraph5\_aMC@NLO (MG5\_aMC)}
framework~\cite{madgraph} for the calculation of predictions at colliders. We
extend the SM by a composite pseudo-scalar $a$
that exhibits small couplings to the SM fermions, gauge bosons, and the Higgs
boson, and is a singlet under the SM symmetries. The pseudo-scalar is modelled as
having a mass of less than 100~GeV, and we consider a parametrisation in which
the couplings and mass are independent. In practice, we augment the SM
Lagrangian with the effective Lagrangian
${\cal L}_a$~\cite{Cacciapaglia:2017iws},
\begin{equation}
  \mathcal{L}_a = \frac12\left(\partial_\mu a\right)\left(\partial^\mu a\right)
    - \frac{1}{2}M_{ a }^2 a ^2 - \sum_f \frac{iC_fm_f}{f_{ a }} a \bar{\Psi}_f\gamma^5\Psi_f +
\frac{g_s^2\kappa_g a }{16\pi^2f_{ a }}G^{a}_{\mu\nu}\tilde{G}^{a\mu\nu}  + \frac{g^2\kappa_W a }{16\pi^2f_{ a }}W^i_{\mu\nu}\tilde{W}^{i\mu\nu} + \frac{g^{\prime 2}\kappa_B a }{16\pi^2f_{ a }}B_{\mu\nu}\tilde{B}^{\mu\nu},
\label{eq:efflag}
\end{equation}
where $M_a$ is the mass of the pseudo-scalar and the sum indicates a sum over
all SM fermion fields $\Psi_f$ with masses $m_f$. The $C_f$ and $\kappa_V$ (with
$V=g, W, B$)
parameters are model-specific and control the couplings of $a$ to fermions
and gauge bosons respectively. The $C_f$ coefficient is universal for the lighter fermions whose masses arise from four-fermion interactions. Although the coupling for the top, $C_t$, may take several values depending on the representation of the top partner in the partial compositeness mechanism~\cite{Cacciapaglia:2017iws}, we have followed the convention of Ref.~\cite{Cacciapaglia:2017iws} and taken $C_f = C_t$. For the 12 benchmark scenarios under
consideration, we present the coefficients dictating the coupling of the
pseudo-scalar in Table~\ref{tab:coeffs}.

\begin{table}
 \renewcommand\arraystretch{1.55}
 \setlength\tabcolsep{8pt}
\begin{tabular}{c||c|c|c|c|c|c|c|c|c|c|c|c}
& M1 & M2 & M3 & M4 & M5 & M6 & M7 & M8 & M9 & M10 & M11 & M12 \\
\hline
\hline
$\kappa_g$ & -7.2 & -8.7 & -6.3 & -11. & -4.9 & -4.9 & -8.7 & -1.6 & -10 & -9.4 & -3.3 & -4.1 \\
\hline
$\kappa_W$ & 7.6 & 12. & 8.7 & 12. & 3.6 & 4.4 & 13. & 1.9 & 5.6 & 5.6 & 3.3 & 4.6\\
\hline
$\kappa_B$ & 2.8 & 5.9 & -8.2 & -17. & .40 & 1.1 & 7.3 & -2.3 & -22. & -19. & -5.5 & -6.3 \\
\hline
$C_f$ & 2.2 & 2.6 & 2.2 & 1.5 & 1.5 & 1.5 & 2.6 & 1.9 & .70 & .70 & 1.7 & 1.8 \\
\end{tabular}
  \caption{Relevant couplings~\cite{Cacciapaglia:2017iws} in the twelve
    models~\cite{Belyaev:2016ftv} of interest used as benchmarks in this study.
    We distinguish the coefficients controlling the couplings of the
    pseudo-scalar to the SM gauge bosons ($\kappa_V$) and fermions ($C_f$).}
\label{tab:coeffs}
\end{table}

 In our notation, $g_s$, $g$ and $g'$ refer to the
strong, weak and hypercharge coupling constants, and $G_{\mu\nu}$, $W_{\mu\nu}$
and $B_{\mu\nu}$ ($\tilde G_{\mu\nu}$, $\tilde W_{\mu\nu}$ and $\tilde
B_{\mu\nu}$) stand for the associated (dual) field strength tensors. The decay
constant $f_a$ of the pseudo-scalar $a$, that drives the strength of the
pseudo-scalar couplings to the SM particles, is defined as
\begin{equation}
f_a = \sqrt{\frac{q_\psi^2N_\psi f_\psi^2 + q_\chi^2N_\chi f_\chi^2}{q_\psi^2 + q_\chi^2}},
\end{equation}
which we set to 1~TeV in this analysis, as previous studies~\cite{Cacciapaglia:2017iws} show that the lower bound on $f_a$ is always below 1~TeV for the models
under consideration. This description being effective, we recall that we can
only rely upon it for energies or momenta below a cut-off scale $\Lambda \sim
4\pi f_a$.

%%%%%%%%%%%%%%%%%%%%%%%%%%%%%%%%%%%%%%%%%%%%

\subsection{Pseudo-scalar couplings to gauge bosons}\label{sec:anomcouplings}
\begin{figure}
\includegraphics[width=0.8\textwidth]{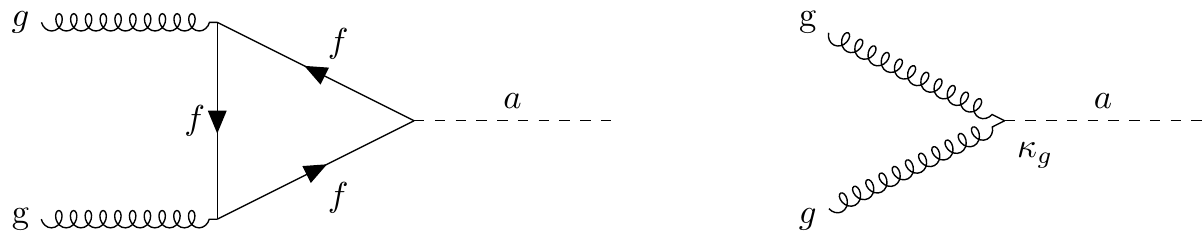}
\caption{The SM (left) and BSM (right) components of the pseudo-scalar coupling
  to gluons, relevant for its production at hadron colliders. The BSM vertex
  consists of an effective Wess-Zumino-Witten structure, as shown in
  Eq.~\eqref{eq:efflag}.}
\label{fig:ggffeyn}
\end{figure}

Leading-order couplings of the form $aVV^\prime$, where $V, V^\prime$ stand for
gauge bosons which may or may not be different, proceed via the
Wess-Zumino-Witten anomaly and are depicted through an effective vertex in the
Lagrangian of Eq.~\eqref{eq:efflag}. However, an SM component where the vertex
is constructed from loops of SM fermions, an example of which is shown in
Fig.~\ref{fig:ggffeyn} (left), is generally significant and should be included.
In order to access each gauge-pseudo-scalar vertex, we rewrite the
gauge-boson interaction part ${\cal L}_V$ of the Lagrangian of
Eq.~\eqref{eq:efflag} in terms of the physical gauge bosons,
\begin{equation}
\mathcal{L}_{V} = \frac{a}{16\pi^2f_a}\left( g_s^2 \kappa_g G_{\mu\nu}\tilde{G}^{\mu\nu} + g^2\kappa_W W^+_{\mu\nu}\tilde{W}^{-\mu\nu} + \right. \\
\left. e^2\kappa_{\gamma\gamma}F_{\mu\nu}\tilde{F}^{\mu\nu} + \frac{e^2\kappa_{ZZ}}{s_W^2c_W^2}Z_{\mu\nu}\tilde{Z}^{\mu\nu}  + \frac{2e^2\kappa_{Z\gamma}}{s_Wc_W}F_{\mu\nu}\tilde{Z}^{\mu\nu}\right) \ ,
\end{equation}
where $c_W$ and $s_W$ denote the cosine and sine of the EW mixing angle, and $e$
is the electromagnetic coupling constant. While the anomalous couplings read
\begin{equation}
\kappa_{\gamma\gamma} = \kappa_W +\kappa_B, \qquad
\kappa_{Z\gamma} = c_W^2\kappa_W - s_W^2\kappa_B \qquad\text{and}\qquad
\kappa_{ZZ} = c_W^4\kappa_W + s_W^4\kappa_B\ ,
\end{equation}
where contributions originating from the SM fermion loops should additionally be
included for all existing interactions ($gga$, $\gamma\gamma a$, $ZZa$,
$W^+W^-a$ and $Z\gamma a$). It is, however, expected that the role of the leptons
and of the five light flavours of quarks is negligible, their couplings
to the pseudo-scalar being suppressed by the smallness of their masses, as are
the contributions from the electroweak bosons that are suppressed by the heavy
propagators running into the loops. In the following, we however stress the
importance of the bottom quark, whose contributions are in fact not so negligible.

\begin{figure}
  \includegraphics[width=0.5\textwidth]{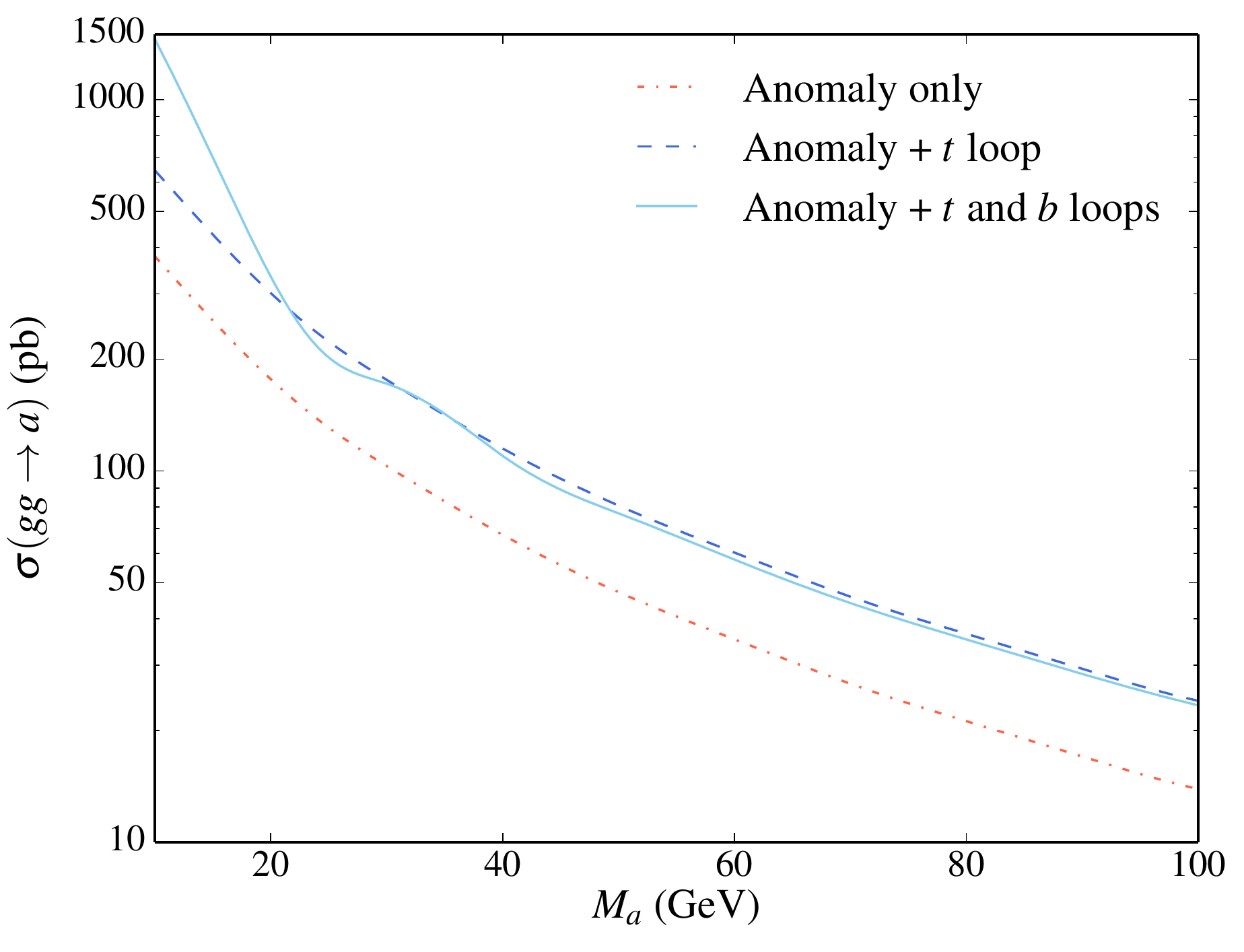}
  \caption{Gluon fusion production cross section for the pseudo-scalar $a$ with an effective gluon-pseudo-scalar coupling of 1 for $M_a \in [10, 100]$~GeV. We distinguish the
    contributions arising solely from the anomalous interactions (dash-dotted
    red), the sum of the latter with the top quark loop contributions (dashed
    blue) and the full predictions including the bottom loop contributions as
    well (solid teal).\label{fig:ggfprod}}
\end{figure}

As an example, we focus on the $gga$ vertex and calculate the
partonic gluon-fusion production cross section of a pseudo-scalar,
\begin{equation}
 \sigma_0 = \frac{1}{256\pi f_a^2} \frac{g_s^4}{16\pi^2}\
    \Big|\kappa_g + \sum_f C_f^2 A(\tau_f)\Big|^2
   \qquad\text{with}\qquad
   \tau_f = \frac{4m_f^2}{M_a^2} \ .
\label{eq:s0}\end{equation}
Such an expression includes the anomaly contribution, as well as the sum over
the contributions from each fermion species. 
The function $A(\tau)$ is defined,
for a given fermion, by
\begin{equation}
  A(\tau) = \tau
  \begin{cases}
                                   -\frac{1}{4}\left[\log\frac{1+\sqrt{1-\tau}}{1-\sqrt{1-\tau}} - i \pi\right]^2 & \text{if}~\tau<1\ , \\[.25cm]
                                   \arcsin^2\left(\frac{1}{\sqrt{\tau}}\right) & \text{if}~\tau\geq 1\ ,\\
  \end{cases}
\label{eq:newf}
\end{equation}
which results from the three-point scalar function of the quark loop propagator.
In the case of top quarks, $\tau_t\geq 1$ and $A(\tau_t)$ is approximately
constant ($\approx 1$) throughout the pseudo-scalar mass range. This thus leads
to an approximately constant increase in the gluon fusion production cross
section relative to the pure anomalous component, as illustrated in
Fig.~\ref{fig:ggfprod}. In the latter, we convolute the expression of
Eq.~\eqref{eq:s0} with the leading order set of NNPDF 2.3 parton densities
\texttt{NNPDF23\_lo\_as\_0130\_qed}~\cite{Ball:2013hta} with $M_a \in [10, 100]$~GeV. In this analysis we ignore lower
masses, where a light pseudo-scalar is subject to strong experimental
bounds~\cite{Cacciapaglia:2017iws}.
The behaviour is quite different in the case of $b$ quarks, where logarithmic
effects produce an oscillation in the contribution to the cross section (see
Fig.~\ref{fig:ggfprod}). At low masses of $a$, the bottom
quark contribution substantially increases the cross section. For higher masses,
there is a small destructive interference between the top and bottom
contributions, leading to a slight decrease in cross section. This shape of the
bottom quark contribution arises due to the form of the three point scalar
function $A(\tau)$ for small $\tau$, where the interplay between the real and
imaginary part of the loop-integral leads to the observed undulation  in the
cross section behaviour. For a detailed investigation into the
corresponding modifications to the pseudo-scalar decay pattern due to the
inclusion of the $b$-quark loop, we refer the reader to Ref.~\cite{Mason:2019zlh}.

We do not include the contributions of any quarks lighter than the $b$ to run in the fermion loop, as the considered mass range of the pseudo-scalar leads to lighter quarks having a negligible impact on the cross section. This can be supported by the observation that the $c$ quark, the next heaviest quark with $M_c=1.275$~GeV, would impact the cross section at most by the amount that the $b$ quark contributes at $M_a\approx 33$~GeV, due to the form of $A(\tau_c)$ for the small $\tau_c$ value. Given that the bottom quark contribution to the cross section is already negligible at this point, all other quark contributions can therefore be ignored.
As a consequence, we will consider throughout this work all bottom and
top quark loop-contributions to the $aVV$ couplings for all SM gauge bosons $V$,
additionally to the tree-level BSM contributions.

%%%%%%%%%%%%%%%%%%%%%%%%%%%%%%%%%%%%%%%%%%%%
%  Section 3: Low mass searches at lepton colliders

\section{A low mass pseudo-scalar at lepton colliders}\label{sec:lowmasssearches}
In this section we estimate the prospects of future $e^+e^-$ collider searches
for this additional light scalar as an alternative to searches at hadronic
colliders. We
are interested, in particular, in designing an analysis that addresses the
parameter space region in which $M_a \in [10, 60]$~GeV, where constraints on
possible light scalars are particularly weak~\cite{Cacciapaglia:2017iws}. This
mass window has indeed been left relatively open due to few direct searches
performed thus far, the dominant constraint originating from searches for
novel SM Higgs decay modes ($h\to a a$).

We begin this section with a study of pseudo-scalar production at lepton
colliders, which differs from that at hadron colliders as the pseudo-scalar is
produced predominantly in association with other states. We consider a variety of future $e^+e^-$ colliders operating at different c.m.~energies and, in the case of linear colliders, different polarisation options. We then focus on a
circular electron-positron collider aiming to operate at the $Z$-pole and
investigate a possible search channel targeting pseudo-scalar decays into a pair
of tau leptons.

%%%%%%%%%%%%%%%%%%%%%%%%%%%%%%%%%%%%%%%%%%%%

\subsection{Pseudo-scalar production at lepton colliders}\label{sec:prod}

\begin{figure}
\includegraphics[width=0.9\textwidth]{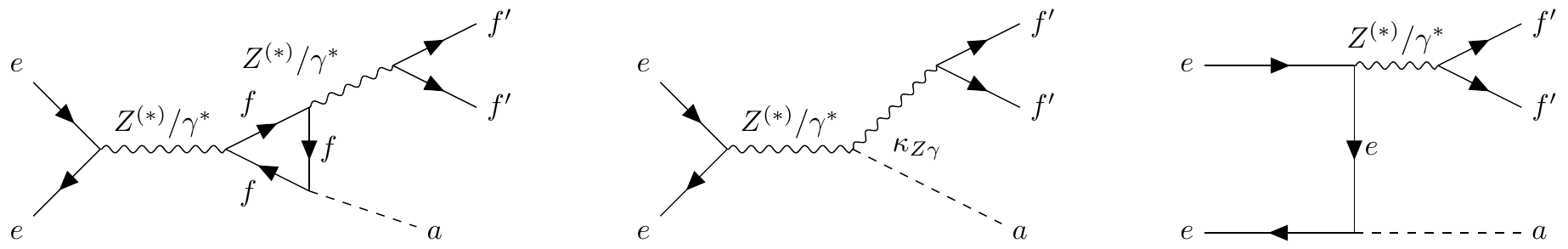}\\[.2cm]
\includegraphics[width=0.6\textwidth]{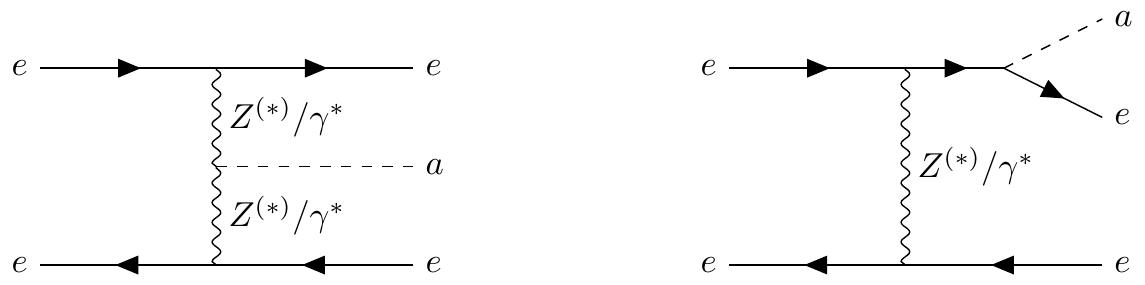}
\caption{Representative Feynman diagrams depicting the production of a
pseudo-scalar in association with a virtual photon, or with a (virtual or real)
$Z$-boson. We include the latter decay into a fermion-antifermion pair (first
line) and additionally consider extra non-resonant diagrams relevant for a $V$
decays into an $e^+e^-$ pair (second line). The first two diagrams denote 
the SM and BSM components of the same process.
% \lara{The production of the pseudo-scalar therefore includes contributions from the $Z$ and $\gamma$ channels, as well as the interference between them}. 
\label{fig:schan}}
\end{figure}

We consider the production of the pseudo-scalar in association with a virtual
photon or with a (virtual or real) $Z$-boson,
%{\lara{should we change this to: We consider the production of the
% pseudo-scalar in association with a (virtual or real) photon or $Z$-boson}
% \benj{no: the photon has to be virtual}}
that we generically denote as $V$,
and that `decays' into any pair of fermions. This mainly proceeds via the
Feynman diagrams shown in the first row of Fig.~\ref{fig:schan}, where one
distinguishes a tree-level contribution in which an $e^+e^-$ pair annihilates
into a $Va$ system through the Wess-Zumino-Witten anomalous $\kappa_{\gamma Z}$
coupling (central diagram), a loop-induced contribution (left diagram) where SM
top and bottom quarks are running in the loop (see
Sec.~\ref{sec:anomcouplings}), and a $t$-channel contribution (right diagram).
In the case where the $V$-boson leads to an $e^+e^-$ final state, extra
non-resonant diagrams additionally contribute (second row of
Fig.~\ref{fig:schan}). We identify two potentially appealing signatures that
differ by the considered final state: the production of the pseudo-scalar
in association either with a pair of opposite-sign first or second generation
leptons $\ell=e,\mu$, or with a pair of light jets $j$ ({\it i.e.}~not
originating from the fragmentation of a $b$-quark),
\begin{equation}
  e^+ e^- \to\ell^+\ell^-a\ , \qquad\qquad e^+e^-\to jja \ .
\label{eq:processes}\end{equation}
Across all models, the branching ratios of the pseudo-scalar $a$ into $\tau^+
\tau^-$ and $b\bar b$ pairs are the highest, as the coupling of the
pseudo-scalar to fermions is proportional to the corresponding fermion mass.
For further insights into the branching modes of the pseudo-scalar under
the influence of the bottom-quark loop effects, the reader is referred to
Ref.~\cite{Mason:2019zlh}, where the rankings of the branching ratios are
demonstrated. Across the models considered, the $b\bar{b}$ and $\tau^+\tau^-$
branching ratios are consistently the most promising avenues.
We therefore choose to design an analysis
dedicated to probing those decay modes, as they are not only the most
abundant, but also feature final-state objects not too difficult to
reconstruct, and offer several handles to extract a signal from the background,
as will be shown below.

The future colliders currently under study include both linear
(ILC~\cite{Behnke:2013xla} and CLIC~\cite{Charles:2018vfv})
and circular (FCC-ee~\cite{Abada:2019zxq} and
CEPC~\cite{CEPCStudyGroup:2018ghi}) possibilities. Electron-positron colliders
may provide a promising avenue through which to search for a light
pseudo-scalar. Many future lepton colliders indeed offer low c.m.~energies
associated with high luminosities, which may allow for the detection of weakly
interacting light particles. While linear colliders offer the benefit of beam polarisation, where some processes such as $W$-boson fusion Higgs production (which depend on the chirality of the colliding leptons) may be amplified, circular colliders allow for the accumulation of higher luminosities, useful when searching for particles that couple timidly or are not copiously produced.
They moreover offer the possibility of hadron collider upgrades.

\begin{figure}
  \includegraphics[width=.48\columnwidth]{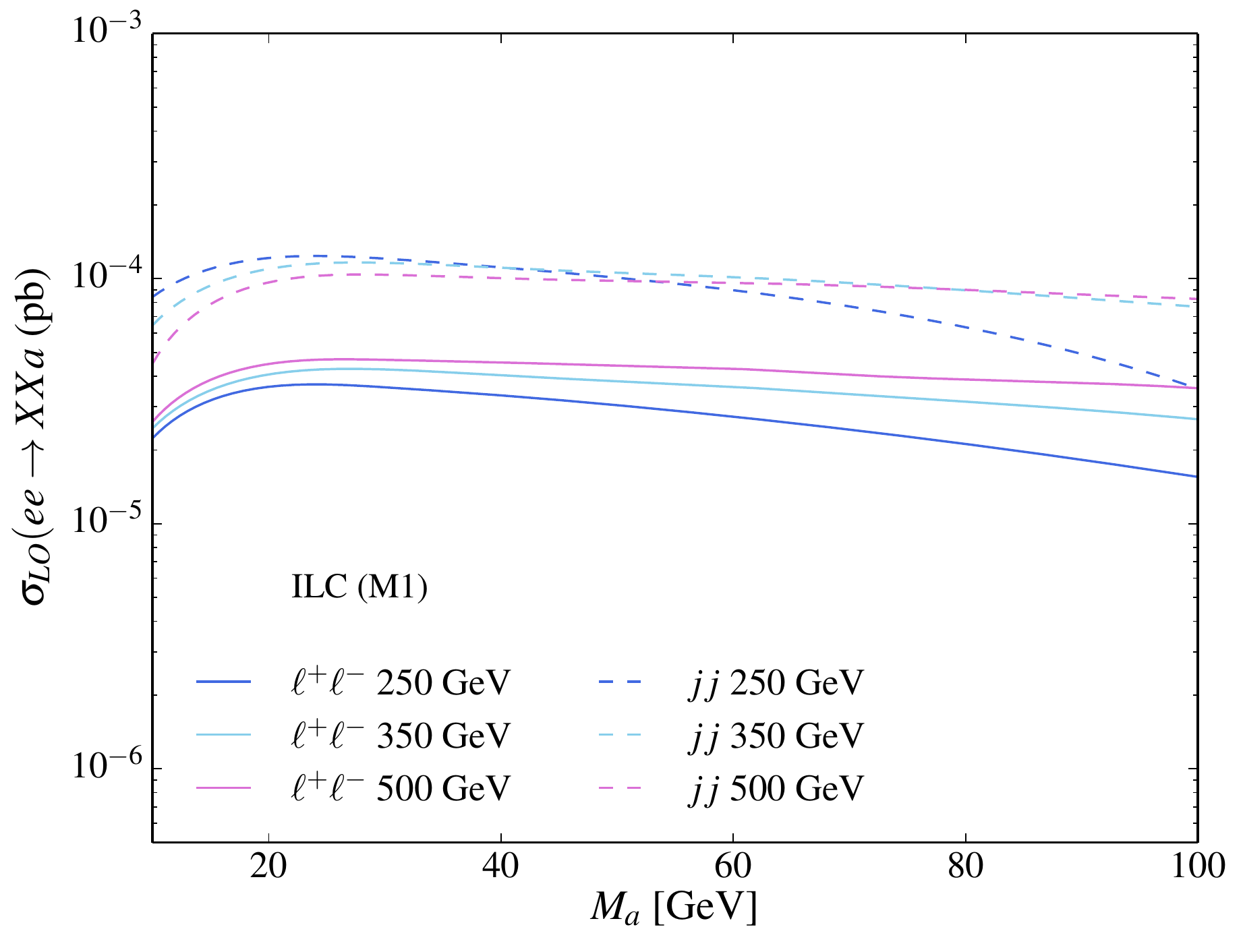}
  \includegraphics[width=.48\columnwidth]{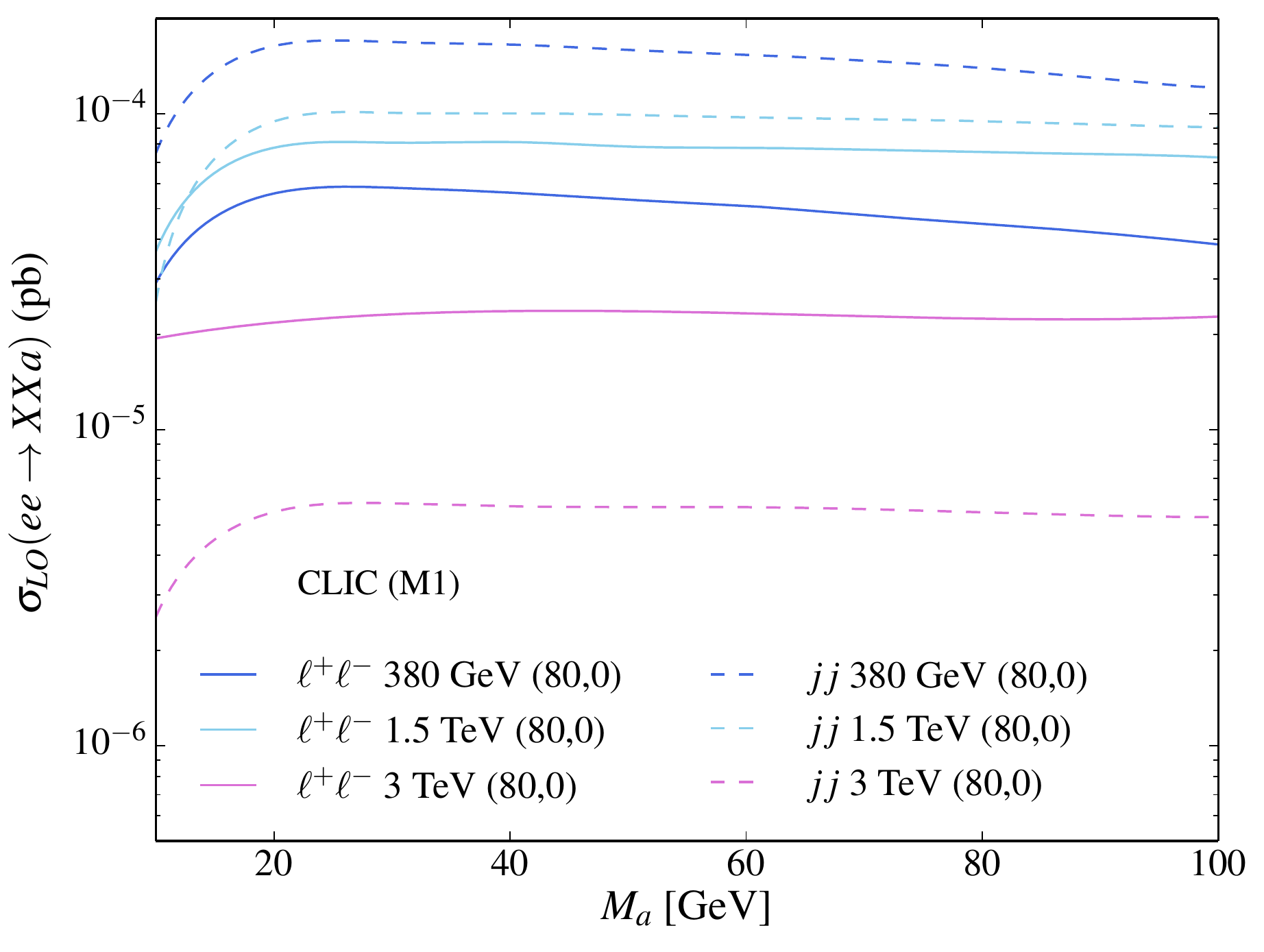}
  \caption{Total cross section associated with the production of the
    pseudo-scalar $a$ in association with leptons (solid lines) or jets (dashed
    lines) at the ILC and CLIC, for various c.m.~energies 
    and including beam polarisations where relevant.\label{fig:ILCCLIC}}
\end{figure}

\begin{figure}
  \includegraphics[width=0.7\textwidth]{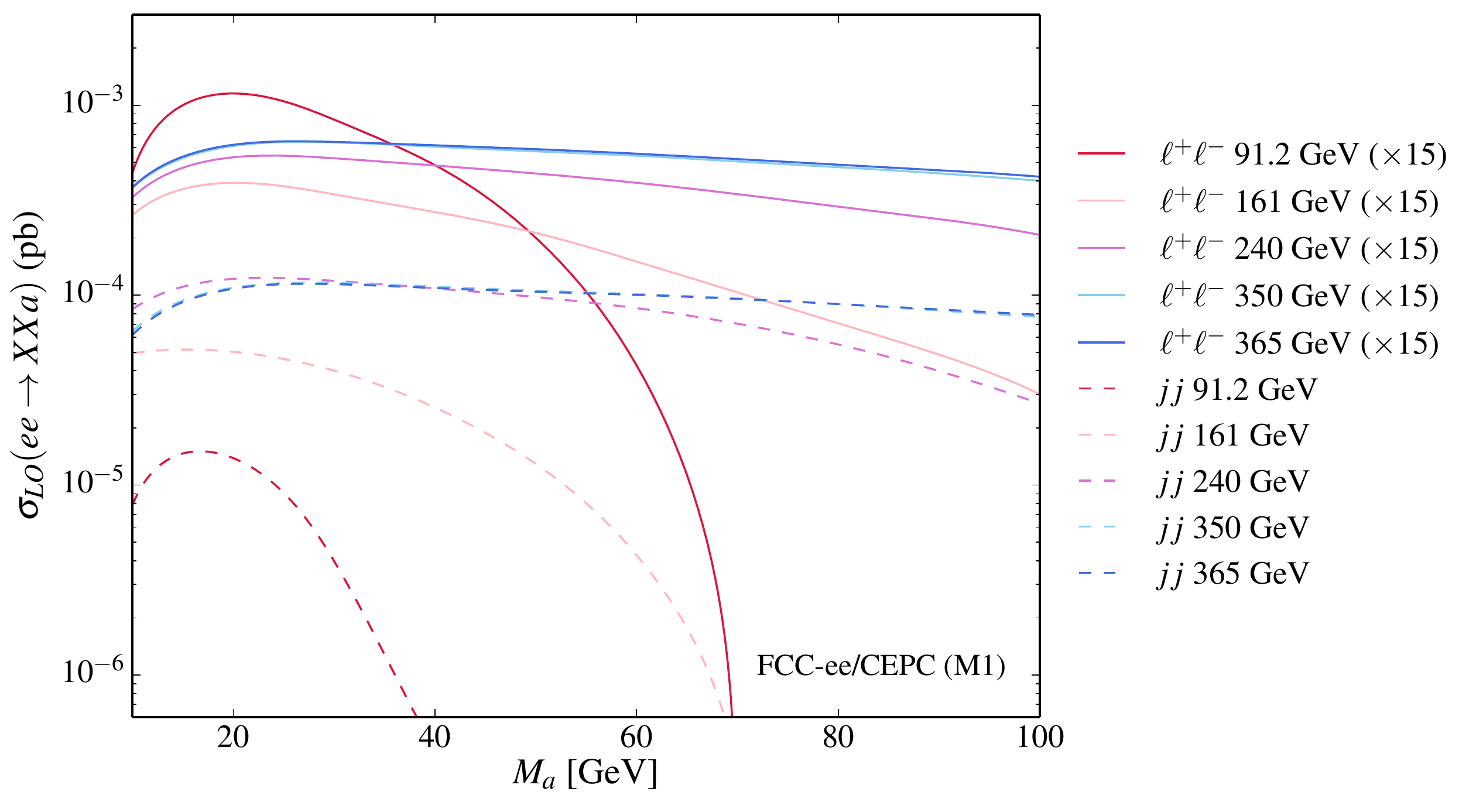}
  \caption{Same as in Fig.~\ref{fig:ILCCLIC}, but for the FCC-ee and CEPC
    circular colliders. \label{fig:FCCCEPC}}
\end{figure}

We first proceed with the calculation of the production cross section associated
with all processes of interest, for each of the aforementioned future lepton
collider options and the varied c.m.~energy choices that have been proposed in
the literature. Using {\sc MG5\_aMC}, we hence present in
Fig.~\ref{fig:ILCCLIC} total cross sections for the two
processes of Eq.~\eqref{eq:processes} for the ILC and CLIC linear colliders. We
rely, as a benchmark, on the model M1 and show leading-order (LO) predictions
for the production of the pseudo-scalar $a$ in association with leptons (solid
lines) and jets (dashed lines). We also considered a variety of beam polarisations relevant for the ILC collider, however, there was no appreciable change in behaviour for different polarisations. In Fig.~\ref{fig:FCCCEPC} we focus instead on the
CEPC and FCC-ee circular colliders, our predictions for pseudo-scalar production
in association with leptons being multiplied by a factor of
15 for legibility. Our results include basic selections on the final-state
leptons and jets, their transverse momentum $p_T$ and pseudo-rapidity $\eta$
being enforced to satisfy
\begin{equation}
  p_T(j)    > 20~{\rm GeV} \ , \quad |\eta(j)| < 5 \ ;\qquad
  p_T(\ell) > 5~{\rm GeV} \ ,  \quad |\eta(\ell)| < 2.5 \ .
\end{equation}
Moreover, jets and leptons are required to be well separated in the transverse
plane, by an angular distance $\Delta R$ of at least 0.4,
\begin{equation}
  \Delta R(\ell,\ell') > 0.4 \ , \qquad
  \Delta R(j,j') > 0.4 \ .
\end{equation}
From these figures we are able to gain an understanding of the potential
pseudo-scalar abundance at future lepton colliders.

Though linear colliders enjoy a relatively constant production cross section
of about 0.01--0.1~fb throughout the entire probed pseudo-scalar mass range
($M_a < 100$~GeV), circular colliders, operating at lower c.m.~energies, are
subject to a fall-off of production cross section at higher masses as the
available phase space decreases. However, for the parameter space region in
which we are interested (featuring $M_a < 60$~GeV), the production rates lie in
the same ballpark regardless of the collider under consideration. As linear
colliders are
subject to much lower expected integrated luminosities, we focus, in the
following analysis outline, on a circular collider. Their larger expected
luminosities may prove crucial in a search for weakly coupled particles, while
operating at relatively low c.m.~energies is useful in reducing otherwise
significant backgrounds, such as originating from $t\bar{t}$ and di-boson
processes.

\subsection{A case study at a circular collider operating at the $Z$-pole}\label{sec:fccee}

We propose an analysis at a future high luminosity lepton collider aiming at
operating at the $Z$-pole and an integrated luminosity of 150~ab$^{-1}$, which
corresponds to the FCC-ee expectation. While the signal production cross section
is subject to a steep decline for higher masses of the pseudo-scalar $a$, the
mass range of interest ($M_a \in [10, 60]$~GeV) is still relatively well covered,
as shown in the previous subsection (see Fig.~\ref{fig:FCCCEPC}).

As a choice for the detector parametrisation,
we consider the IDEA detector concept of the FCC-ee project. This detector, as
any detector project at any future circular electron-positron machine, is
designed using recent technological advances to take advantage of the
exceptionally large data samples due to be delivered by the forecast integrated
luminosities. IDEA is planned to be constructed with a short drift wire chamber
and calorimeter, and includes a low mass superconducting solenoid coil. The
drift chamber will allow for high precision momentum measurements and
good tracking capabilities, as well as excellent particle identification
performance through cluster counting when combined with the dual readout
calorimeter~\cite{Abada:2019zxq}. In particular, this aims at achieving a much
improved impact parameter resolution over that of LEP, as well as a better
momentum resolution and electromagnetic calorimeter resolution, and a finer
electromagnetic calorimeter transverse granularity~\cite{Dam:2018rfz}. Precise
measurements of charged objects properties at lower energies are therefore
clearly achievable.

In building our analysis, we will moreover only consider the channel in which
the pseudo-scalar $a$ is produced in association with a pair of opposite-charge
leptons. This allows us to avoid the difficulty in dealing with the multi-jet
background still reasonably present at lepton colliders. Moreover, we will focus
on the case where the pseudo-scalar decays into a pair of hadronic tau leptons
for two reasons.
% \lara{to remove: First, the $b$-tagging performance is known to be quite poor in
% a regime yielding soft objects (originating from the decay of a light $a$ boson
%into $b$-quarks).}
% Next, h%
Hadronic tau decays account for approximately 2/3 of all
tau decays, and future $e^+e^-$ collider detectors are expected to have
excellent handles on the associated decay vertices. This hence allows for a very
efficient tau reconstruction. In addition, events with
electrons or muons in the final state are expected to be reconstructed with a
very good resolution~\cite{Yan:2016xyx}. Basing our estimations on previous
detectors, we hence expect a typical systematic uncertainty on lepton
identification of around 1\%, together with a systematic uncertainty on hadronic
tau identification of around 2--5\% for taus with a transverse momentum
$p_T>20$~GeV, and up to 15\% otherwise~\cite{Pingel:2224284}.

\begin{figure}
  \includegraphics[width=0.48\columnwidth]{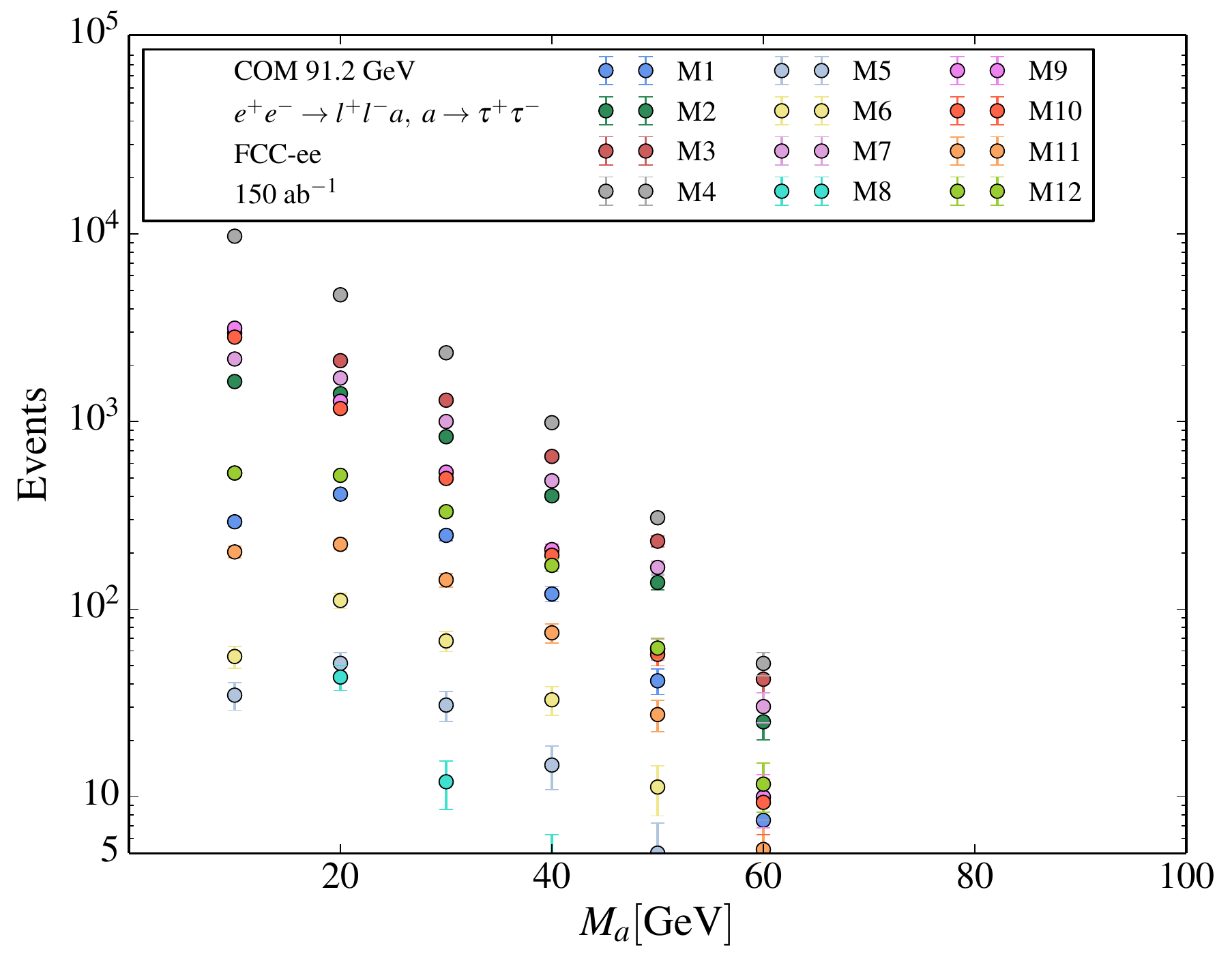}
  \caption{Expected number of $\ell\ell\tau\tau$ signal events in 150~ab$^{-1}$
    of $e^+e^-$ collisions at the $Z$-pole, for the 12 benchmark models under
    consideration in this work. \label{fig:FCClla1}}
\end{figure}

In Fig.~\ref{fig:FCClla1} we display the number of signal events expected from
the production of the pseudo-scalar $a$ in association with a pair of oppositely
charged leptons, with a subsequent pseudo-scalar decay into a pair of hadronic
tau leptons,
\begin{equation}
  e^+ e^- \to \ell^+ \ell^- a \to \ell^+ \ell^- \tau^+ \tau^- \ .
\end{equation}
We consider the entire mass range of interest and include
statistical error bars. This illustrates the motivation to expect a significant
number of new physics events in electron-positron collisions at the $Z$-pole. It
moreover attests that the sensitivity of the machine will depend on the exact
details of the model, as there is a significant range in the expected number of
new physics events across the considered models.

We now design our analysis from Monte Carlo simulations of
both the signal and background processes, using {\sc MG5\_aMC} employed in
conjunction with {\sc Pythia}~8~\cite{Sjostrand:2007gs} to describe parton
showering (which includes both QED and QCD initial state radiation
effects) and hadronisation. The IDEA detector response has been simulated by
relying on the {\sc Delphes}~3~\cite{deFavereau:2013fsa} software package, that
makes use of the anti-$k_T$ algorithm~\cite{Cacciari:2008gp} as implemented in
{\sc FastJet}~3~\cite{Cacciari:2011ma} for event reconstruction. Both these last
codes are driven through their interface with the {\sc MadAnalysis}~5 platform~%
\cite{Conte:2012fm,Conte:2018vmg}, that we also use to carry on our
phenomenological analysis. We begin with a cut-and-count analysis,
aiming at unravelling the signal from the overwhelming backgrounds. However, as a
consequence of the low statistical significance, we then employ a novel machine
learning algorithm based on boosted decision trees in an attempt to improve the
significance, using the {\sc XGBoost} toolkit~\cite{Chen:2016btl}.

In identifying backgrounds to the signal process, we consider both processes
which lead to a true $\ell\ell\tau\tau$ final state, and those containing fakes.
Background events of the first category feature prompt taus that are most likely
to arise from the production of an $\ell^+\ell^-$ pair in association with a
pair of opposite-sign tau leptons through one or two (virtual) $Z$ bosons and
photons. Given the relatively low c.m.~energy of 91.2~GeV, potential background
events originating from two virtual weak bosons or top quark decays are not
expected to contribute much. Background events of the second category arise from
jets faking hadronic taus. They appear in processes such as vector boson
production in association with a pair of jets, where the boson then decays to a
pair of leptons, as well as in processes where the intermediate boson is virtual
or the mediation occurs through virtual photons,
\begin{equation}
  e^+e^- \rightarrow j j \ell^+\ell^- \ .
\end{equation}
In the following, we refer to those background as $Z\text{+jets}$,
although contributions from virtual photons and their interferences are accounted for as well. The
corresponding cross section is $6.65\times 10^{-4}$~pb, that should then be
multiplied by the appropriate fake rate factor. The IDEA detector
parametrisation shipped with {\sc Delphes} includes a 0.1\% misidentification
rate for jets faking hadronic taus and a tau identification efficiency of 60\%~\cite{IDEAsim}. This tau identification efficiency is based on efficiencies at existing colliders, and is expected to be a conservative estimate. We therefore expect a fake contribution of the order of $10^{-10}$~pb, which can
thus safely be neglected.

%%%%%%%%%%%%%%%%%%%%%%%%%%%%%%%%%%%%%%%%%%%%

\subsection{A cut-and-count analysis}\label{sec:cutcount}
Our selection closely follows the pattern of the final state under
consideration. We require events to contain at least two leptons ($N_\ell \geq
2$), each with a minimum $p_T$ of 10~GeV to ensure good reconstruction, and at
least two hadronic taus ($N_\tau\geq 2$), each with $p_T>5$~GeV. We moreover
enforce a minimum invariant mass $m_{\ell\ell}>12$~GeV on the lepton pair
produced in association with the di-tau system, which is necessary to eliminate
non-prompt leptons and
avoid low mass hadronic resonances. Similarly, the invariant mass of the tau pair
$M_{\tau\tau}$ is constrained to be at least 10~GeV. Given that the hadronic tau
decay mode of the pseudo-scalar leads to neutrinos which carry away momentum,
the di-tau invariant mass spectrum is expected to be soft and peak below the
pseudo-scalar mass $M_a$. It is therefore useful to maintain low momentum
thresholds for the tau pair where possible. To summarise, our preselection
imposes that
\begin{equation}
  N_\ell \geq 2 \ \ \text{with}\ \ p_T(\ell) > 10~{\rm GeV};\quad
  N_\tau \geq 2 \ \ \text{with}\ \ p_T(\tau) >  5~{\rm GeV};\quad
  M_{\ell\ell} > 12~{\rm GeV};\quad
  M_{\tau\tau} > 10~{\rm GeV}.
\label{eq:preselection}\end{equation}
After this preselection, we expect about 50,000 background events for signal
event counts ranging from below 1 ($M_a \lesssim 10$~GeV) or a few
($M_a\gtrsim 50$~GeV) to 10--40 (10~GeV $ < M_a < $ 50~GeV).

\begin{figure}
  \includegraphics[width=0.48\textwidth]{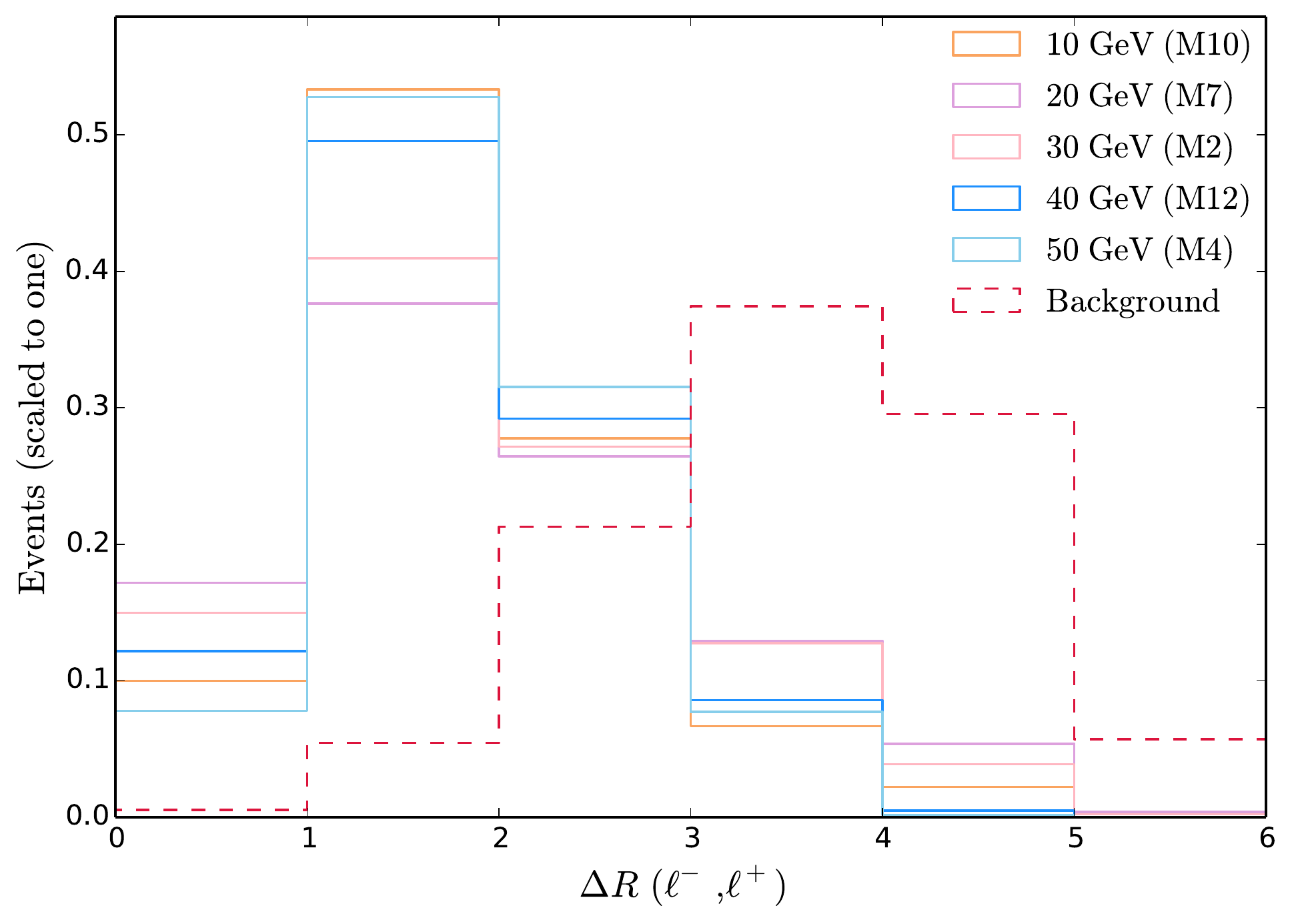}
  \includegraphics[width=0.48\textwidth]{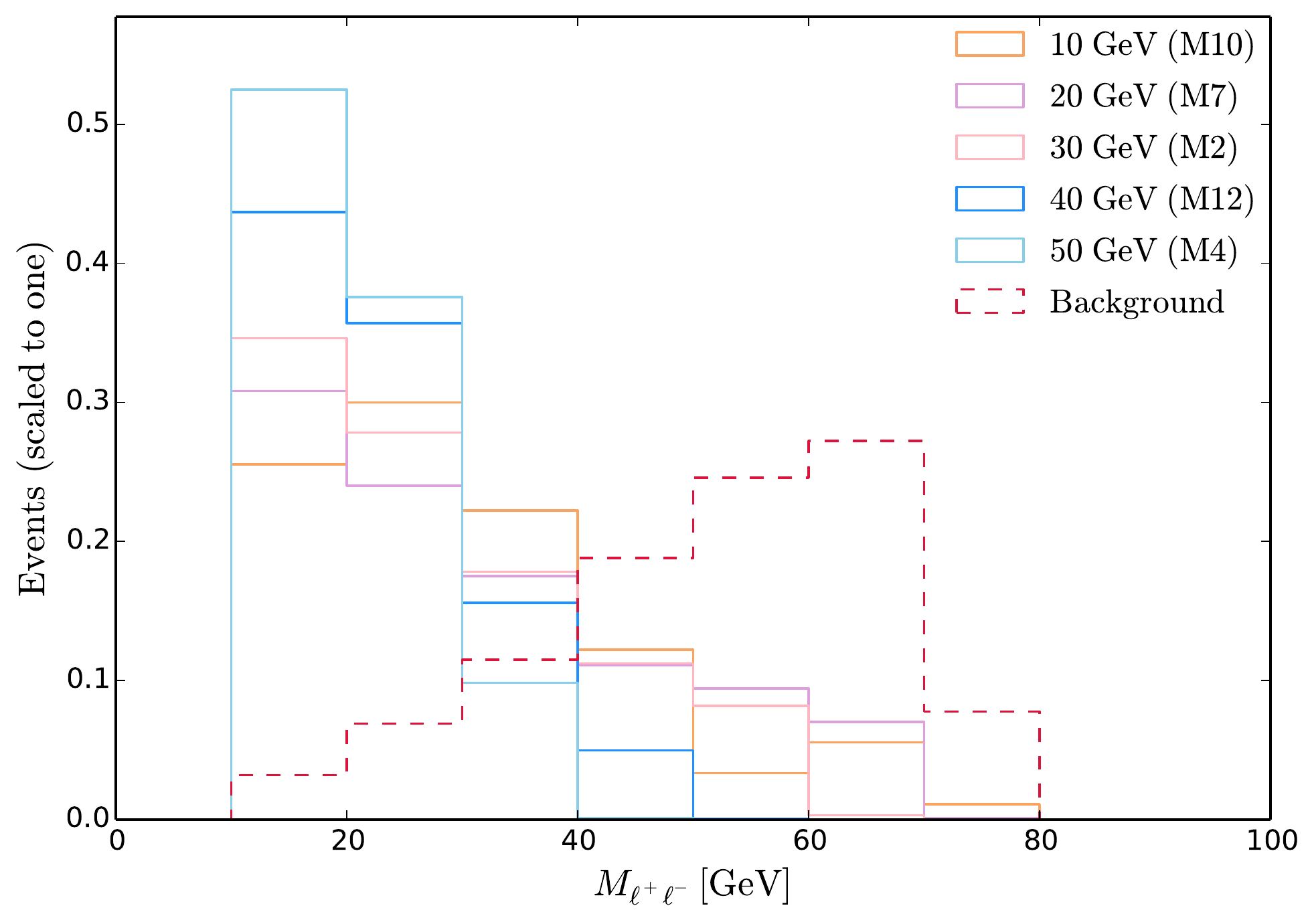}
  \caption{Properties of the di-lepton system for signal and background events
    originating from electron-positron collisions at a c.m.~energy of 91.2~GeV.
    We consider various representative signal hypotheses, and focus on the
    angular distance in the transverse plane between the two leptons (left), and
    their invariant-mass spectrum (right). \label{fig:za}}
\end{figure}

In order to reject the background while keeping a large signal efficiency, we
investigate first the properties of the two final-state leptons. As in the case
of the signal they are produced together with a low mass resonance, we expect
the presence of potentially discriminating features in various kinematic
distributions, the exact details behind those features being related to the
resonance mass. We present, in Fig.~\ref{fig:za}, the angular separation between
the two leptons $\Delta R(\ell^+, \ell^-)$ (left) as well as the di-lepton
invariant mass distribution $M_{\ell\ell}$ (right). We consider both the
$Z$+jets background (red dashed) as well as five signal hypotheses from
different models and pseudo-scalar masses below 60~GeV ({\it i.e.}~our
pseudo-scalar mass range of interest). More precisely, we have chosen a
selection of five models (M2, M4, M7, M10 and M12) exhibiting a variety
of hypercolour group and coset structures (see Tab.~\ref{tab:models}). This
allows us to largely explore the possibilities of the considered class of
composite scenarios. The results depicted in the figures demonstrate that the
$\Delta R (\ell^+,\ell^-)$ and $M_{\ell\ell}$ spectra tend to peak at higher
values for the background than for the illustrative signal hypotheses. This
suggests two interesting cuts to isolate the pseudo-scalar signal,
\begin{equation}
  \Delta R (\ell^+,\ell^-) < 3\ ; \qquad\qquad
  M_{\ell\ell} < 40~{\rm GeV}.
\label{eq:lepton_cuts}\end{equation}

\begin{figure}
  \includegraphics[width=0.48\textwidth]{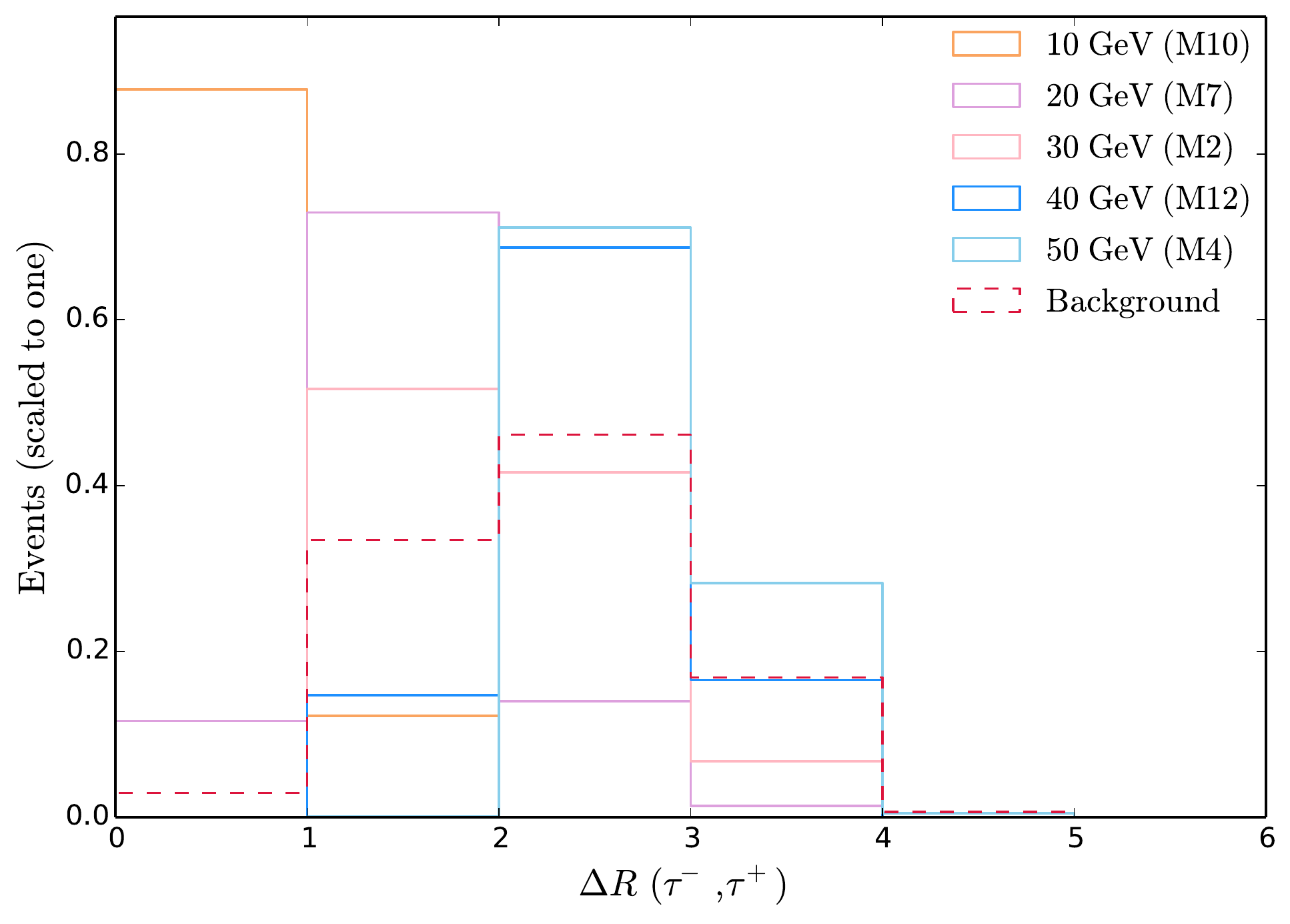}
  \includegraphics[width=0.48\textwidth]{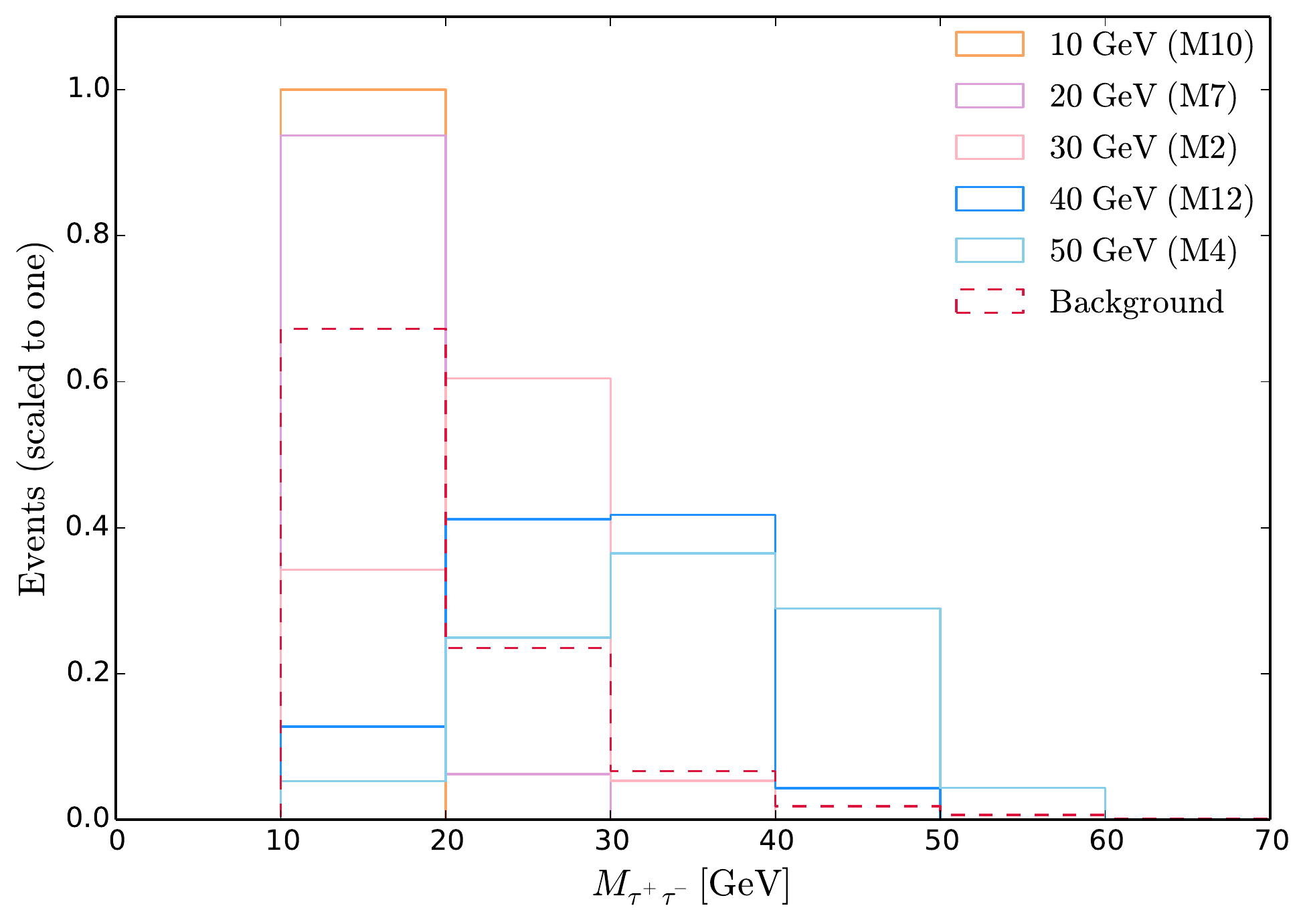}
   \caption{Same as in Fig.~\ref{fig:za} but for the di-tau system and after the
    cuts of Eq.~\eqref{eq:lepton_cuts}.\label{fig:mtatabkgsig}}
\end{figure}

In addition, we also make use of the properties of the di-tau system to extract
our composite signal from the background. In the signal case, the pair of
hadronic tau leptons is issued from the decay of a resonance, so that its
properties are expected to be largely different from the background case. In
particular, the invariant mass of the di-tau system is expected to peak at a
value just below the mass of the the pseudo-scalar, as a result of the presence
of the neutrinos originating from the tau decays and carrying away some
momentum~\cite{Bartschi:2019xlg}. The resulting distribution is shown in the
left panel of Fig.~\ref{fig:mtatabkgsig}. As expected, the invariant mass of the
di-tau system is shifted relative to the mass of the pseudo-scalar, and peaks
just below the true value of the latter. The background distribution is, however,
not so well differentiated from the signal one, as the di-tau system features
predominantly a low invariant mass, as driven by the selection cuts of
Eq.~\eqref{eq:lepton_cuts}. We nevertheless define five signal regions, each of
them being dedicated to one specific pseudo-scalar mass hypothesis, and
respectively impose
\begin{equation}
  M_{\tau\tau} < 10, 20, 30, 40, 50~{\rm GeV}.
\end{equation}
This allows us to eliminate some background in the heavier pseudo-scalar cases.
The minimum requirement on $M_{\tau\tau}$ of Eq.~\eqref{eq:preselection} that
protects us from the contamination of QCD resonances makes us, however, unable to
get further handles on the signal for pseudo-scalar masses of 10~GeV or smaller.
Similarly, we investigate the potential of the angular separation between the
two taus (right panel of Fig.~\ref{fig:mtatabkgsig}). Although shape differences
are visible, they do not allow for a clear separation of the signal and the
background. Any related cut will therefore be omitted from our analysis. 

\begin{table}
  \setlength\tabcolsep{8pt}
  \def\arraystretch{1.3}
  \begin{tabular}{c|c c c c c }
  Model & $M_a=$ 10 GeV & $M_a=$ 20 GeV & $M_a=$ 30 GeV & $M_a=$ 40 GeV & $M_a=$ 50 GeV\\
  \hline
  M2   & 0.0015  &  0.13  &  0.090  & 0.049&   0.020 \\
  M4   &  0.0013 &  0.42  &   0.26   & 0.12  &  0.040  \\
  M7   &  0.0024  &  0.14 &   0.11   & 0.061 &   0.023 \\
  M10   &  0.0042   & 0.11  &   0.055  & 0.023 &    0.0078\\
  M12    &  0.00061 &  0.047 &  0.035 & 0.021 & 0.017  \\
  \end{tabular}
  \caption{Sensitivity of our cut-and-count analysis expressed as the
    significance $S/\sqrt{S+B}$ for a selection of considered composite
    scenarios and 150~ab$^{-1}$ of electron-positron collisions at the
    $Z$-pole.}
  \label{tab:ccresul}
\end{table}

In Tab.~\ref{tab:ccresul} we present the expected sensitivity of our
cut-and-count analysis in terms of standard deviations defined by an
$S/\sqrt{S+B}$ figure of merit, $S$ and $B$ respectively representing the
number of selected signal and background events. We find that, given the
relative rareness of the signal events amongst an abundance of background, it is
difficult to obtain any hope to observe even a $1\sigma$ deviation from the
background-only hypothesis across the entire mass range considered. It is, however, possible that this could be ameliorated by designing more appropriate and dedicated variables like the missing mass. As reflected
in the $M_a$-dependence of the cross section shown in Fig.~\ref{fig:FCCCEPC}, the
significance is maximised at $M_a = 20$~GeV. For larger pseudo-scalar masses,
the steep fall-off of the cross section indeed reduces $S$ to too large a
level. 

%%%%%%%%%%%%%%%%%%%%%%%%%%%%%%%%%%%%%%%%%%%%

\subsection{A machine-learning-based analysis}\label{sec:ml}
In order to improve the figure of merit of our analysis, we move on with
considering a machine learning algorithm. We rely on the {\sc XGBoost}
toolkit~\cite{Chen:2016btl} that allows for utilising gradient
boosted tree methods~\cite{friedman2001} while offering fast training speed
coupled with a good accuracy~\cite{10.5555/2996850.2996854}.

In general, a
machine learning algorithm employing a tree ensemble uses a series of additive
optimisations computed from a given set of variables to predict an output,
{\it i.e.}~in our case the classification of an event as a signal or a
background event. At each stage of the training process, gradient boosting
modifies the existing constraints in order to correct the classification errors
made by the current best set of optimisations, continuing until no further
improvement can be made in considering the residuals and errors of the prior
stages. The {\sc XGBoost} toolkit includes a novel algorithm geared towards the
handling of sparse data, which is useful in our case as both signal and
background events may not fully populate the event space.

The performance of the algorithm for a given set of optimisations can be
evaluated by a quantity denoted as the area under the curve (\textit{auc}). This
corresponds to the integral of the receiver operating characteristic (\textit{roc})
depicting the dependence of the signal purity of the events selected by the
algorithm, $S/(S+B)$, on the signal selection efficiency $S/S_0$, where $S_0$
stands for the total number of signal events provided to the algorithm. The
{\it auc} metric hence represents the degree of separability between background
and signal. In addition, we use the approximate median discovery significance
(\textit{ams}) to estimate the sensitivity of the analysis to our signal. It is
defined by~\cite{Cowan:2010js}
\begin{equation}
  \textit{ams} = \sqrt{2\left( \left( S + B \right) \ln \left( 1 + \frac{S}{B}\right)-S \right)},
\end{equation}
where $S$ and $B$ can also be seen as the true and false positives respectively.
While the \textit{ams} provides the discovery potential of the analysis, its
usage as an evaluation metric and learning objective is unstable and may lead to
overfitting. The performance of the algorithm was therefore optimised using
the \textit{auc} quantity, following which the corresponding \textit{ams} was
calculated.

\begin{figure}
  \includegraphics[width=0.55\textwidth]{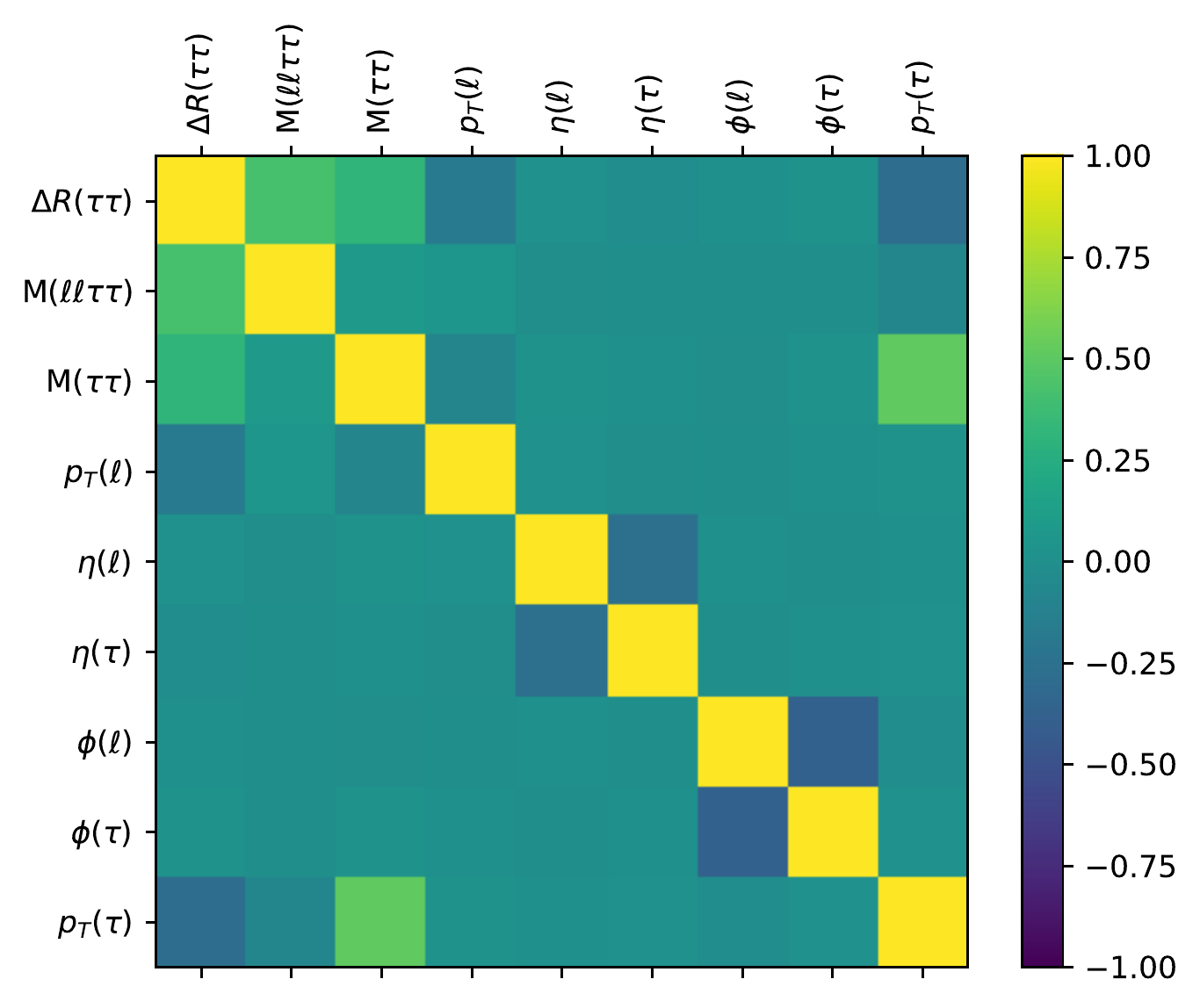}
  \caption{Correlations amongst the nine kinematic variables employed in our
    machine learning exercise. The set of variables includes the angular
    separation of the hadronic tau pair $\Delta R(\tau\tau)$, the invariant mass
    of the combined lepton-tau system $\ell\ell\tau\tau$ as well as of the di-tau
    system, and the transverse momenta $p_T$, pseudo-rapidities $\eta$ and
    azimuthal angle $\phi$ of the taus and leptons.\label{fig:corrplot}}
\end{figure}

After applying the preselection of Eq.~\eqref{eq:preselection}, we derive a set
of uncorrelated kinematic variables to be used as input to our machine learning
algorithm. They consist of a combination of primary variables (the tau and
lepton transverse momenta, pseudo-rapidities and azimuthal angles) and derived
variables (the di-tau invariant mass $M_{\tau\tau}$ and angular separation
$\Delta R(\tau, \tau)$, as well as the invariant mass of the $\ell\ell\tau\tau$
system) that have been chosen such that their importance to the machine learning
algorithm was maximised while removing any variables that were too strongly
correlated with the others. The variables and their correlations are depicted in
Fig.~\ref{fig:corrplot}. The objective of the {\sc XGboost} learning task was
set to a logistic regression for binary classification. At each step (also known
as splitting), the tree booster constructs new classifiers by combining and
weighting the classifiers obtained at the previous step, the initial classifiers
being the input variables. The hyperparameters that were found to affect the
performance of the method were the learning rate, the maximum tree depth and the
minimum child weight. The learning rate controls data over-fitting by varying
the learning step size, the maximum depth of a tree indicates how many times a
tree can split (hence controlling the algorithm complexity), and the minimum
child weight controls the minimum weight that can be assigned when designing a
new classifier.

\begin{table}
  \setlength\tabcolsep{8pt}
  \def\arraystretch{1.3}
  \begin{tabular}{c|c|c c c c c}
    Model & Metric & $M_a=$ 10 GeV & $M_a=$ 20 GeV & $M_a=$ 30 GeV & $M_a=$ 40 GeV & $M_a=$ 50 GeV\\
    \hline
    \multirow{2}{*}{ M2} & \textit{auc} & 0.98$\pm$0.003 & 0.87 $\pm$ 0.006  & 0.84  $\pm$ 0.0013  & 0.94 $\pm$0.0058 &  0.95 $\pm$ 0.0066 \\
&   \textit{ams} &  0.22 & 2.96 & 2.41 &  0.29 & 0.11   \\
    \hline
    \multirow{2}{*}{ M4} &  \textit{auc} & $0.98\pm 0.0045$ & $0.95\pm 0.0029$  & $0.87\pm 0.020$  & $0.88\pm 0.042$ &  0.89$\pm$0.061 \\
    &   \textit{ams} & 1.16  & 2.83 & 1.69 & 0.54  & 0.15   \\
    \hline
    \multirow{2}{*}{ M7}  &  \textit{auc} & $0.98\pm 0.0018$ & $0.86\pm 0.0082$  & $0.88\pm 0.0011$  & $0.90\pm 0.0012$ &  $0.94\pm$ 0.019\\
    &   \textit{ams} & 0.22  & 3.20 & 2.58 & 0.27  & 0.14   \\
    \hline
    \multirow{2}{*}{ M10}&  \textit{auc} & 0.98$\pm$0.003 & 0.92$\pm$ 0.0057 & 0.90$\pm$0.019  & 0.96$\pm$0.0078 &  0.96$\pm$0.0050 \\
    &   \textit{ams} & 0.37  & 4.08 & 2.35 & 0.14  & 0.042   \\
    \hline
    \multirow{2}{*}{ M12} &  \textit{auc} & 0.98$\pm$0.0075 & 0.92$\pm$0.003  & 0.92$\pm$ 0.013 & 0.95$\pm$0.0044 & 0.96 $\pm$0.0082 \\
    &   \textit{ams} & 0.066  & 1.26 & 0.98 & 0.11  & 0.046   \\
  \end{tabular}
  \caption{{\sc XGBoost} evaluation metric and significance obtained for a
    representative set of composite scenarios and pseudo-scalar masses, using
    150~ab$^{-1}$ of electron-positron collisions at the $Z$-pole.
  \label{tab:xgresul}}
\end{table}

In our analysis, 80\% of the available Monte Carlo data was used for training
purposes, and the remaining 20\% for testing. For each model and $M_a$ value, we
tuned the hyperparameters using a $k$-fold cross-validation, so that the choice
maximising the \textit{auc} was adopted. In particular, the maximum depth
parameter was kept low and early stopping was employed in order to control
over-fitting. It was found that a maximum depth of 3, a minimum child weight of
1 and a learning rate of 0.3 gave the most desirable result across the entire
range of considered models. The \textit{auc} metric and the corresponding
significances obtained for a representative set of models are indicated in
Tab.~\ref{tab:xgresul}.

The results indicated in Tab.~\ref{tab:xgresul} display a general improvement
over the traditional cut-and-count method, but also an important variation
across models and pseudo-scalar masses. In particular, the significance peaks at
$M_a$~=~20~GeV for all models, as this corresponds to the maximum of the signal
cross section (see Fig.~\ref{fig:FCClla1}). However, there are large differences
in the trends across the models. For example, the performance for
the model M10 quickly falls to one of the lowest for $M_a$~=~50~GeV. These
behaviours reflect not only the varying production cross sections across the
models but also the variations in the kinematics resulting from differing
Lagrangian parameters. On the other hand, we find a low significance for
$M_a$~=~10~GeV, where despite a relatively large cross section, the preselection
cuts (and in particular the $M_{\tau\tau}$ requirement) rejects a large potion
of the signal.

The best performance of our analysis is found for scenarios featuring $M_a =
20,30$~GeV. For all models, the performance then drops off quickly for $M_a =
40,50$~GeV, and it falls more sharply than it does in the cut-and-count
case. Such a drop in significance at these pseudo-scalar masses is expected, as
the cross section decreases with $M_a$. Moreover, some signal kinematic
distributions exhibit important variations with the pseudo-scalar mass. An
example can be taken from the $\Delta R (\tau^-,\tau^+)$ spectrum (see the right
panel of Fig.~\ref{fig:mtatabkgsig}), where scenarios with higher $a$ masses
($M_a=40,50$~GeV) lead to very similar signal and background distribution
shapes. Finally, we have trained the gradient boosting algorithm using one
hyperparameter choice across all considered masses within a given model, and the choice of kinematic
variables on which to train the models was guided by a focus on the lower mass
setups. This training has been done in isolation for each model in order
to find the hyperparameters best suited in each case. This path has clearly optimised the 20/30~GeV scenarios, as they were
expected to yield the highest significance by virtue of the larger associated
cross sections. The potential price to pay could be a less efficient training
for higher masses of $a$.

In comparing the significance trends of the gradient boosting results with those
of the cut-and-count method, we observe the same ranking of performance among
the different models, with the exception of the model M4. This model corresponds
to the highest cross sections, and may therefore have been expected to be better
performing. However, our framework leads to overfitting for the $M_a=20,30$~GeV
cases, which had to be carefully controlled by using an early stopping of the
algorithm. This resulted in a lower significance.

\begin{table}
  \def\arraystretch{1.3}
  \setlength\tabcolsep{8pt}
  \begin{tabular}{c c| cc | cc}
  \multirow{2}{*}{Model} & \multirow{2}{*}{$M_a$ [GeV]} &
    \multicolumn{2}{c|}{Cut and Count} &  \multicolumn{2}{c}{Machine Learning}\\
  &    &  2$\sigma$ \qquad & 3$\sigma$&  2$\sigma$ & 3$\sigma$ \\
  \hline
  \multirow{5}{*}{M2} &10 &  2.67$\times 10^{8}$ & 6.00$\times 10^{8}$& 1.24$\times 10^{4}$ & 2.79$\times 10^{4}$  \\
                            & 20 &3.55$\times 10^{4}$  & 7.99$\times 10^{4}$ & 68.5 &  154 \\
                             & 30& 7.41$\times 10^{4}$ & 1.67$\times 10^{5}$& 103 & 232 \\
                             & 40& 2.50$\times 10^{5}$ & 5.62$\times 10^{5}$ & 7.13$\times 10^{3}$ & 1.61$\times 10^{4}$\\
                             & 50& 1.50$\times 10^{6}$ & 3.38$\times 10^{6}$ & 4.96$\times 10^{4}$ & 1.12$\times 10^{5}$\\                            
  \hline
  \multirow{5}{*}{M4} &10 & 3.55$\times 10^{8}$ & 7.99$\times 10^{8}$ & 446 &  1.00$\times 10^{3}$\\
                            & 20 & 3.40$\times 10^{3}$ & 7.65$\times 10^{3}$ & 74.9 & 169\\
                            &  30& 8.88$\times 10^{3}$ & 2.00$\times 10^{4}$& 210 & 473 \\
                            &  40& 4.17$\times 10^{4}$ &  9.38$\times 10^{4}$ & 2.06$\times 10^{3}$ & 4.63$\times 10^{3}$\\
                            &50  & 3.75$\times 10^{5}$ & 8.44$\times 10^{5}$ & 2.67$\times 10^{4}$ & 6.00$\times 10^{4}$\\                              
\hline
\multirow{5}{*}{M7} & 10& 1.04$\times 10^{8}$ & 2.34$\times 10^{8}$ & 1.24$\times 10^{4}$ & 2.79$\times 10^{4}$\\
                            & 20 & 3.06$\times 10^{4}$ & 6.89$\times 10^{4}$ & 58.5 & 132 \\
                             & 30&4.96$\times 10^{4}$  & 1.12$\times 10^{5}$ & 90.1 & 203 \\
                             & 40& 1.61$\times 10^{5}$ & 3.63$\times 10^{5}$ & 8.23$\times 10^{3}$ & 1.85$\times 10^{4}$\\
                             & 50 & 1.13$\times 10^{6}$ & 2.55$\times 10^{6}$ &3.06$\times 10^{4}$ & 6.89$\times 10^{4}$\\                              
\hline
\multirow{5}{*}{M10} &10 & 3.40$\times 10^{7}$ & 7.65$\times 10^{7}$ & 4.38$\times 10^{3}$ & 9.86$\times 10^{3}$ \\
                           &  20 & 4.96$\times 10^{4}$ & 1.12$\times 10^{5}$ & 36.0 & 81.1 \\
                           & 30  & 1.98$\times 10^{5}$ & 4.46$\times 10^{5}$ & 109  & 244\\
                           & 40  & 1.13$\times 10^{6}$ & 2.55$\times 10^{6}$ &3.06$\times 10^{4}$ & 6.89$\times 10^{4}$\\
                            & 50 & 9.86$\times 10^{6}$ & 2.22$\times 10^{7}$ & 3.40$\times 10^{5}$ & 7.65$\times 10^{5}$\\                              
\hline
\multirow{5}{*}{M12} & 10& 1.61$\times 10^{9}$  & 3.63$\times 10^{9}$ & 1.38$\times 10^{5}$ & 3.10$\times 10^{5}$\\
                            & 20 & 2.72$\times 10^{5}$ & 6.11$\times 10^{5}$ & 378 & 850 \\
                            &30  & 4.90$\times 10^{5}$ & 1.10$\times 10^{6}$ & 624 & 1.41$\times 10^{3}$\\
                             &40 & 1.36$\times 10^{6}$ & 3.06$\times 10^{6}$ &4.96$\times 10^{4}$ & 1.12$\times 10^{5}$\\
                           & 50 & 2.08$\times 10^{6}$ & 4.67$\times 10^{6}$ &  2.84$\times 10^{5}$ & 6.38$\times 10^{5}$\\                              
\end{tabular}
  \caption{Required luminosities, in ab$^{-1}$, to obtain a $2\sigma$ and
    $3\sigma$ significance to the pseudo-scalar signal at a future
    electron-positron collider operating at the $Z$-pole. We present results for
    our cut-and-count (third and fourth columns) and gradient boosting (fifth
    and sixth columns) methods, for an illustrative selection of models.}
\label{tab:reqlum}
\end{table}

In Tab.~\ref{tab:reqlum}, we translate our results in terms of the luminosity
that is needed in order to achieve a significance of 2$\sigma$ (to preclude the
existence of the new resonance) or 3$\sigma$ (to claim evidence for the
resonance) at a future electron-positron collider aiming at operating at the
$Z$-pole. The table also shows the gain obtained by using the gradient boosting
algorithm over a more traditional cut-and-count method. Very importantly, our
findings show that for certain models, an achievable integrated luminosity would
yield a $2\sigma$ or even 3$\sigma$ significance. In all cases, larger
pseudo-scalar masses remain likely out of reach at a c.m.~energy of 91.2~GeV, as
does the $M_a~=~10$~GeV case.

%%%%%%%%%%%%%%%%%%%%%%%%%%%%%%%%%%%%%%%%%%%%
%  Section 5: Conclusion

\section{Conclusion}\label{sec:conclusion}
In this work we have designed an analysis targeting a light pseudo-scalar,
ubiquitous to composite Higgs models, at a future electron-positron collider
aiming at collecting a large luminosity at the $Z$-pole. In our predictions we
have considered the pseudo-scalar couplings to gauge bosons to full leading
order, {\it i.e.}~by including relevant effects stemming from loops of $b$
quarks. The latter have a significant impact for low mass pseudo-scalars, unlike
what is traditionally assumed, and should be considered both at present and
future hadron and lepton colliders.

We have demonstrated the possibility of actually getting hints for a low mass
pseudo-scalar at a future lepton collider operating at a centre-of-mass energy
of 91.2~GeV, focusing on the production mode in which the pseudo-scalar is
produced in association with a pair of electrons or muons and decays into a pair
of hadronic taus. The corresponding Standard Model background has been found
difficult to reduce via a standard cut-and-count analysis, which resulted in a
poor sensitivity and a rare signal entirely hidden within the large background.
In an attempt to improve these findings we have made use of a machine learning
algorithm based on boosted decision trees. It yielded an improvement in
sensitivity in almost all cases. In particular, we have observed a marked
improvement for scenarios in which the pseudo-scalar mass $M_a = {20,30}$~GeV,
where the related significance approaches $3\sigma$ at an integrated luminosity
of 150~ab$^{-1}$. Lighter configurations ($M_a\lesssim10$~GeV) are not
promising, given that the signal is expected to be dominated by the background
and mostly annihilated by any decent event preselection. The significance also
drops off for higher pseudo-scalar masses, by virtue of a decreasing signal
cross section and key kinematic properties becoming very similar to the
background ones. It is, however, possible that this could be ameliorated by
designing more appropriate and dedicated variables.
Additionally, while this analysis has focused on the four-lepton final state, a further work may investigate the $e^+e^-\rightarrow a\gamma$ channel which may be competitive, where similar studies~\cite{Bauer:2018uxu} have shown the $e^+e^-\rightarrow a\gamma(a\rightarrow \ell\ell)$ signature to be more constraining for axion-like particles. In focusing on the four lepton final state we have shown that, using machine learning capabilities, even weaker signals are accessible.

We have used, as an example, the framework of a future electron-positron collider aiming at operating at the $Z$-pole and collecting a very large integrated luminosity. In this context, we have demonstrated that search avenues for weakly coupled particles are promising. The analysis that we have proposed complements earlier works, such as in Ref.~\cite{Cacciapaglia:2017iws}, focusing on potential options for the LHC, and fills a gap for what concerns light states. It has indeed been shown that there is a scarcity of searches for such a light pseudo-scalar in the mass region that has been considered in this paper~\cite{Cacciapaglia:2019bqz}. In particular, the Higgs to BSM branching ratio often provides the sole (indirect) constraint in the mass region $M_a \in [10, 60]$~GeV~\cite{Cacciapaglia:2017iws, Cacciapaglia:2019bqz}. Additionally, investigations into future collider prospects for light pseudo-scalars have been previously presented in Refs.~\cite{Bauer:2018uxu, Bauer:2017ris}. In those broad comprehensive reviews of existing bounds on axion-like particles, our channel is complementary to those studied, and falls within the bands of unconstrained parameter space. The proposed analysis would be complementary to existing di-photon searches~\cite{Mariotti:2017vtv}, where a drop in sensitivity in the di-photon channel corresponds to increased sensitivity within the di-tau channel~\cite{Cacciapaglia:2017iws} (and vice versa) for the models M1-12 across the mass range here considered. Additionally, this analysis could reach below the existing 90~GeV lower bound for di-tau searches~\cite{CMS:2016pkt}. In this we find our pseudo-scalar to be a candidate which evades all existing bounds, and present this channel as an additional one which may be accessed through machine learning capabilities.

 From our findings, we demonstrated that a direct search for a light composite
pseudo-scalar at high integrated luminosity lepton colliders should be seriously
considered. While our generic analysis covers the parameter space region in
which the mass of the pseudo-scalar is less than $60$~GeV, it is certainly less
sensitive to $M_a$ values of 40~GeV or more. Future works should determine
whether it could be optimised for these heavier configurations, perhaps by
considering future lepton colliders operating at higher centre-of-mass energies.
Among the avenues to be explored, one could benefit from a gain in sensitivity
by relying on the spin-0 nature of the pseudo-scalar and assessing the potential
of various angular distributions between pairs of final-state objects. For the
same reason, it may also be useful to make use of
tau polarisation in order to separate signal from the
background~\cite{Assamagan:2001jw}. Finally, other options may rely on the
presence of a second heavier pseudo-scalar $\eta^\prime$, that is common
to many composite models.

%%%%%%%%%%%%%%%%%%%%%%%%%%%%%%%%%%%%%%%%%%%%
%  Acknowledgements

\section*{Acknowledgements}
We thank Giacomo Cacciapaglia, Gabriele Ferretti, Thomas Flacke,  Michelangelo Mangano and Andrea Thamm for useful discussions. ASC is supported in part by the National Research Foundation of South Africa (NRF) and thanks the University of Lyon 1 and IP2I for support during the collaboration visit in Lyon. LM is supported by the UJ GES 4IR initiative, and thanks Campus France and SA-CERN for additional support. The authors would like to acknowledge the Mainz Institute for Theoretical Physics (MITP) of the Cluster of Excellence PRISMA+ (Project ID 39083149) for its hospitality and support.

\bibliographystyle{utphys}
\bibliography{phdbib}

\end{document}